\documentclass[eqsecnum,amsmath,preprintnumbers,superscriptaddress,nofootinbib,aps,11pt]{revtex4}%,twocolumn,10pt]
\pdfoutput=1
\usepackage{tikz}
\usepackage{bm}
\usepackage{MnSymbol}
\usepackage{amsmath,accents}
\usepackage{graphicx}
\usepackage{hyperref}
\usepackage{bbold}
\usepackage{eufrak}
\usepackage{upgreek}
\usepackage{stmaryrd,scalerel}
\usepackage{enumitem}
\usepackage{color} %enables use of coloured text
\usepackage{etex}
\usepackage{morefloats}

 \usepackage{longtable}


\DeclareFontFamily{U}{matha}{\hyphenchar\font45}
\DeclareFontShape{U}{matha}{m}{n}{
      <5> <6> <7> <8> <9> <10> gen * matha
      <10.95> matha10 <12> <14.4> <17.28> <20.74> <24.88> matha12
      }{}
\DeclareSymbolFont{matha}{U}{matha}{m}{n}
\DeclareMathSymbol{\varrightharpoonup}{3}{matha}{"E1}

\def\beq{\begin{equation}}
\def\eeq{\end{equation}}

\def\bea{\arraycolsep .1em \begin{eqnarray}}
\def\eea{\end{eqnarray}}
\def\Tr{{\rm Tr}}

\def\a{\alpha}
\def\b{\beta}
\def\A{ \mathcal{A}}

\def\nn{ \nonumber \\}

\newcommand{\pp}{{\hat{\phi}}}
\newcommand{\cc}{{\hat{\chi}}}

\def\eq#1{(\ref{#1})}

\def\0#1#2{\frac{#1}{#2}}

\def\grgl{\:\hbox to -0.2pt{\lower2.5pt\hbox{$\sim$}\hss}{\raise3pt\hbox{$>$}}\:}
\def\klgl{\:\hbox to -0.2pt{\lower2.5pt\hbox{$\sim$}\hss}{\raise3pt\hbox{$<$}}\:}

\begin{document}
\title{Equivalence of effective actions}

\author{Kevin Falls}
\address{\footnotesize\mbox{Instituto de F\'isica, Facultad de Ingenier\'ia, Universidad de la Rep\'ublica, J.H.y Reissig 565, 11300 Montevideo, Uruguay}}

\date{\today}

\begin{abstract}
For the same quantum field theory distinct effective actions can be obtained by coupling sources to different choices of field variables. This is the same as considering effective actions for theories related by a change of variables and thus differ only by the values of so called inessential couplings. The effective actions will appear quite different since they generate correlation functions of different operators. Here we show that the effective actions are related by an implicit change of variables of the mean field, i.e. the argument of the effective action. Conversely, one can go the other way: by making a change of variables of the mean variable we  obtain a new effective action which generates correlation functions of an implicitly defined composite field. The existence of the implicit transformations in both cases rests on the existence of solutions to initial value problems where the ``time'' parameter is an inessential coupling. Non-perturbatively the solutions may only exist for some non-zero amount of this time. However, at each order in perturbation theory one obtains linear equations which implies that unique solutions exist.  We then show that scattering amplitudes are independent of inessential couplings without use of perturbation theory. For gauge theories we expect our correspondence to extend to effective actions which differ by a choice of gauge.
\end{abstract}

\maketitle

\newpage
\tableofcontents

\newpage

\section{Introduction}
Suppose we have two sets of correlations functions for the same quantum field theory (QFT) where one set is for the field variable $\hat{\chi}$, in terms of which we express the action $S[\cc]$, and the other set is for another field variable $\hat{\phi}$.  If the two choices of field variables are related by invertible maps
\beq \label{hat_cov}
\cc = \cc[\pp]  \,, \,\,\,\,\,\,  \pp  = \pp[\cc]\,, 
\eeq
the correlation functions will contain the same information. Here we will have in mind suitably local maps such that $\cc(x) =\cc^x[\pp]$ is a function for the field $\pp(x)$ and its derivatives.  
The two sets of correlation functions can be packaged inside effective actions   $\Gamma_{\hat{\chi}}[\chi]$ and $\Gamma_{\hat{\phi}}[\phi]$, i.e. the generating functionals of one-particle irreducible correlation functions. The arguments of the effective actions, $\phi$ and $\chi$, are the respective mean field values, $\phi = \langle \pp\rangle$ and $\chi  = \langle \cc \rangle$,  in the presence of sources which couple linearly to the respective field variables.
 Ignoring quantum effects the effective actions are approximated by classical actions such that 
\beq \label{Gamma_pp_tree}
\Gamma_\pp[\phi] = S_\pp[\phi] + O(\hbar)  = S[\cc[\phi]] + O(\hbar)\,.  
\eeq
On the other hand, including quantum corrections, the effective actions will have rather different forms since they generate different correlation functions.
 In particular, for a non-linear transformation \eq{hat_cov}, 
 \beq
 \Gamma_{\hat{\phi}}[\phi] \neq \Gamma_{\cc}[\cc[\phi]]\,,
 \eeq
 beyond the tree-level approximation \eq{Gamma_pp_tree}. Furthermore to make sense of \eq{hat_cov} a UV cutoff at some finite momentum scale 
 $\Lambda$ needs to be in place.
 For a generic change of variables \eq{hat_cov},  even if $\Gamma_{\hat{\chi}}[\chi]$ is renormalised, and so finite as $\Lambda \to \infty$, generally $\Gamma_{\hat{\phi}}[\phi]$ will  be UV divergent.
 In order that both effective actions are finite, as $\Lambda \to \infty$, will require us to renormalise the relation \eq{hat_cov}.
 Thus, as is the case for $S[\cc]$, generally the renormalised relation will be ill defined as we remove the cutoff in the limit $\Lambda \to \infty$.

Despite these complications, given the immense freedom to make a change of variables, one suspects that particular choice of the field variable can simplify the calculation of a particular observable of interest. However, guessing the form of this transformation is likely to be challenging given the need to calculate quantum corrections to the transformation law \eq{Gamma_pp_tree}.

Seemingly independently of  \eq{hat_cov}, one can consider  expressing  $\Gamma_{\hat{\chi}}[\chi]$ as a functional of a different mean field variable $\phi$, such that
\beq \label{mean_cov}
\chi = \chi[\phi]   \,, \,\,\,\,\,\,   \phi  = \phi[\chi] \,,
\eeq
are invertible maps. In this case writing the action $\Gamma_{\hat{\chi}}'[\phi]=\Gamma_{\hat{\chi}}[\chi[\phi]]$ amounts to merely  expressing the effective action in terms of a new variable $\phi$. It is important to stress that \eq{mean_cov} is {\it not} a change of field variables, since $\chi$ is not the field but the {\it expectation value of the field} in the presence of a source. The only true change of variables are the are the {\it genuine} transformations \eq{hat_cov} between the quantum field operators.  Nonetheless, it is very tempting to transform the effective action in via  \eq{mean_cov} since the effective action contains all quantum corrections and the issue of renormalising a composite operator does not arise. One might hope therefore that  \eq{mean_cov} corresponds to an {\it effective} change of variables induced by some genuine changes of variables.
The purpose of this paper is to explore this possible correspondence.

Considering linear transformations $\pp  = \int {\rm d}^dy \beta_1(x ,y) \cc(y) + \beta_0(x) $
there is a trivial correspondence  between the two kinds of transformations since in this case $\pp[\langle \cc \rangle] = \langle \pp \rangle$ and hence $\phi[\chi] = \pp[\chi]$ is the corresponding effective change of variables. 
Here we will argue that there is a non-trivial correspondence  between  genuine transformations \eq{hat_cov}  and effective transformations \eq{mean_cov} in case of non-linear transformations. Specifically, the correspondence we shall uncover implies that 
 for each genuine  transformation there is a dual effective transformation and visa-versa such that 
 \beq \label{GammaequalsGamma}
\Gamma_{\hat{\phi}[\cc]}[\phi] = \Gamma_{\cc}[\chi[\phi]]\,,
\eeq
to all orders in perturbation theory.
%\begin{enumerate}[label=(\roman*)]
%
%\item
%\end{enumerate}
Thus starting from a change of variables \eq{hat_cov}  an effective change of variables \eq{mean_cov} exists such that \eq{GammaequalsGamma} holds. Furthermore, effective actions \eq{GammaequalsGamma} related to $\Gamma_{\hat{\chi}}[\chi]$ by an effective change of variables of the mean field \eq{mean_cov} are related by a genuine transformation \eq{hat_cov} and hence contain the same physical information.

To put this into context, suppose   $\Gamma_{\hat{\phi}}[\phi]$ is arrived at by purely theoretical principles while experimental data instead finds correlation functions of some ``observed field'' $\hat{\chi}$ from which we can construct the phenomenological effective action $\Gamma_{\hat{\chi}}[\chi]$. Then, according to the correspondence, the experimental data is confirming the theory provided we can find  an effective transformation  \eq{mean_cov} relating the effective actions. Thus one can confirm theories by making a change of variables at the level of the effective action without knowing the corresponding change of variables at the level of quantum field operators. The correspondence therefore grants us the freedom to make effective changes of variables when testing a model. Equally, if two different theoretical models lead to effective actions related by effective transformations, we can conclude that the models are equivalent.

Inspiration for this work comes from \cite{Cohen:2022uuw,Cohen:2023ekv} where it was show that if a relation between effective actions of the form \eq{GammaequalsGamma}  
holds then the scattering amplitudes computed from both effective actions will agree. It was then shown at one-loop that identity  \eq{GammaequalsGamma} indeed holds for known changes of variables between $\cc$ and $\pp$.
Similarly, at one-loop it is straight forward to go the other way around, by starting with the relation 
between the mean fields $\chi$ and $\phi$ and inferring the relation between $\hat{\chi}$ and  $\hat{\phi}$.
 Since the equivalence theorem \cite{Chisholm:1961tha,Kamefuchi:1961sb,Efimov:1972juh,Kallosh:1972ap,Arzt:1993gz,Tyutin:2000ht} states that the amplitudes will agree to all loop orders, this suggests the one-loop result can be extended.  

The aim of this paper is not to find the explicit changes of variables beyond one-loop order, but instead to demonstrate their existence implicitly. In particular, we will show that the changes of variables can be defined as the solutions to initial value problems. These initial value problems can be formulated non-perturbatively but require a unique solution to exist. On the other hand, at each order in perturbation theory we obtain a linear equation which allows us to demonstrate the correspondence to each order.

For simplicity we will work explicitly with a single component scalar field theory but the results apply also to more general theories.
They are particularly powerful when applied to a fundamental theory, e.g. of quantum gravity, where the natural microscopic description of a ``quantum spacetime'' may differ substantially from a natural phenomenological  understanding, e.g. in terms of a spacetime metric and the fields of the standard model. In cosmology they allow for directly moving from the Jordon frame and Einstein Frame at the level of the full effective action including loop corrections \cite{Falls:2018olk}.

 In \cite{Wetterich:2024uub} the two types of transformations \eq{hat_cov} and  \eq{mean_cov} have been contrasted and dubbed ``microscopic'' and ``macroscopic''.
According to the correspondence they are in fact two ways of carrying out the same operation. Furthermore, it provides the interpretation of  $\Gamma_{\pp}[\phi] = \Gamma_{\hat{\chi}}[\chi[\phi]]$: while it does not generally generate correlation functions of $\phi[\cc]$ it does generate correlation functions of some quantum operator  $\pp[\cc]$ related by the correspondence to the effective transformation $\phi[\chi]$.

\section{Preliminaries}
One starting point to study a QFT or statistical field theory is to consider a bare action or Hamiltonian $S[\cc]$. Our work has applications both in real time QFT and in statistical field theory so it will be important to consider both cases.
Here we will generally use Euclidean space notation but our considerations will work both for Euclidean and Minkowski space. When working in Minkowski space  $S= -iS^L$ where $S^L$ is the Lorentzian action. 

 To compute correlation functions of the field $\cc$ it is useful to couple a source $J(x)$ to the field $\cc(x)$. In QFT $J(x)$ is usually taken to be purely imaginary whereas it is typically taken to be real when we work in Euclidean space.
From the action\footnote{Here and throughout a dot ``$\cdot$'' implies an integral over spacetime i.e. $J \cdot \cc = \int d^dx J(x) \cc(x)$.} $S[\hat{\chi}]-  J \cdot \cc$ we can obtain the generating functional of connected correlation functions $W_\cc[J]$ which is defined by
\beq
e^{W_\cc[J]}   = \int {\rm d} \hat{\chi}   \, e^{- S[\hat{\chi}]  + J \cdot \cc}\,.
\eeq
Equivalently, the effective action, that generates the one-particle irreducible correlation functions of the  field $\hat{\chi}$, is given by
   \beq \label{GammaSt}
e^{- \Gamma_{\hat{\chi}}[\chi] } = \int {\rm d} \hat{\chi}   \, e^{- S[\hat{\chi}]  + ( \hat{\chi} - \chi) \cdot   \Gamma^{(1)}_\cc[\chi]      }\,.
\eeq
The generating functionals $W_\cc[J]$ and  $\Gamma_{\hat{\chi}}[\chi]$ are related by a Legendre transformation
\beq
\Gamma_{\hat{\chi}}[\chi] = - W_\cc[J]   + J \cdot \chi\,,
\eeq
and thus we can understand $\chi$ as a parameterisation of the source $J = J[\chi]$ where $\chi = \langle \cc \rangle$ is the expectation value of the field $\cc$ in the presence of the source $J$.

 While $\cc$ may appear as a natural variable, we can also consider an alternative effective action for the same theory defined by coupling the source to a field $\pp[\cc]$. In this case we obtain $W_\pp[J]$ defined by 
\beq
e^{W_\pp[J]}   = \int d \hat{\chi}   \, e^{- S[\hat{\chi}]  + J \cdot \pp[\cc]}\,,
\eeq
 which generates connected correlation functions of the field $\pp[\cc]$.
Again the Legendre transformation of $W_\pp[J]$ leads to the effective action $\Gamma_{\hat{\phi}}[\phi]$ which can be defined directly by
\beq \label{Gamma}
e^{- \Gamma_{\hat{\phi}}[\phi] } = \int {\rm d} \hat{\chi}   \, e^{- S[\hat{\chi}]  + ( \hat{\phi}[\hat{\chi}] - \phi) \cdot  \Gamma^{(1)}_\pp[\phi]      }\,.
\eeq
We can understand $\pp$ as a composite operator of the field $\hat{\chi}$, however this interpretation is not unique.
Instead, by a change of integration variables, we can write
\beq \label{Gamma_2nd}
e^{- \Gamma_\pp[\phi] } = \int {\rm d} \hat{\phi}\, e^{- S_\pp[\hat{\phi}]  + ( \hat{\phi} - \phi) \cdot  \Gamma^{(1)}_\pp[\phi]      }\,,
\eeq
where
\beq \label{S_rel}
e^{-S_\pp[\hat{\phi} ] } =  \left|\det \frac{\delta \hat{\chi}[\hat{\phi}] }{\delta \hat{\phi} }  \right|   e^{- S[\hat{\chi}[\hat{\phi}]] }\,.
\eeq
In this way of writing the integral, $\hat{\phi}$ appears to be a fundamental variable for a ``different'' theory. 
 
For a given choice of the field variable $\pp[\cc]$ expectation values of an operator $\mathcal{O}[\cc]$, which depend on $\phi$, are defined by 
\beq  \label{ExpVal}
\left\langle \hat{\mathcal{O}}[\hat{\chi}]\right\rangle_\pp[\phi] :=  e^{ \Gamma_\pp[\phi] }    \int d \hat{\chi}   \, e^{- S[\hat{\chi}]  + ( \hat{\phi}[\hat{\chi}] - \phi) \cdot   \frac{ \delta \Gamma_\pp[\phi]}{\delta \phi}      } \hat{\mathcal{O}}[\hat{\chi}] \,,
\eeq
where the subscript $\pp$ indicates which field the source couples to.
By differentiating \eq{Gamma} with respect to $\phi$ and assuming the invertibility of the hessian of $\Gamma_\pp[\phi]$ (i.e. the existence of a propagator) we obtain 
\beq \label{phi}
\phi = \left\langle \pp[\hat{\chi}]\right\rangle_\pp\,.
\eeq
Thus all expectation values \eq{ExpVal} are functionals of the expectation value of $\pp$ in the presence of a source which is given by
\beq
J_x =  \frac{ \delta \Gamma_\pp}{\delta \phi^x} \,.
\eeq 
Setting $J_x = 0$ all observables, i.e. expectation values $\left\langle \hat{\mathcal{O}}[\hat{\chi}]\right\rangle_\pp$, are independent of the choice of field variable.

Now let's consider making a small change in the form of $\pp[\cc] \to \pp[\cc]  + \delta\pp[\cc]$. From \eq{Gamma} and using \eq{phi} we obtain, to first order in $\delta \pp$,  
\beq \label{Gamma_deltapp}
\Gamma_{\pp+ \delta \pp} = \Gamma_{\pp} -  \left\langle \delta \pp[\hat{\chi}]\right\rangle_\pp \cdot  \frac{\delta \Gamma_{\pp}}{\delta \phi}  + O(\delta \pp^2)\,.
\eeq
This already suggests \eq{GammaequalsGamma} since it has the form of an infinitesimal change of variables for a scalar function. The identity \eq{Gamma_deltapp} implies that small changes in the effective action which vanish on the equations of motion are {\it inessential} since they can be undone by a field transformation. The central  role of \eq{Gamma_deltapp} in demonstrating the equivalence theorem has been stressed in \cite{Tyutin:2000ht} and it will also play  a central role in our considerations here.

Note that identities  such as \eq{S_rel} and  \eq{Gamma_deltapp}  require regularisation. 
 Furthermore,  even if we have renormalised $S[\cc]$ so that $\Gamma_\cc[\chi]$ is finite, generally $\Gamma_\pp[\chi]$ will be divergent (see \cite{Manohar:2024xbh} for a related discussion).  
  Here we will have in mind the framework of effective field theory with a UV cutoff in place.  Then all UV divergences are regularised at some cutoff scale $\Lambda$ which is considered to be part of the effective theory.
We can implement the regularisation in the continuum by  modifying the kinetic term in the action such that the latter has the form 
 \beq \label{Reg_S}
S[\hat{\chi}] =  \frac{1}{2} \int_{x,y} \partial_\mu \hat{\chi}(x) c^{-1}_{\Lambda}(x,y) \partial_\mu \hat{\chi}(y) +   S^{\rm int}[\hat{\chi} ]\,,
 \eeq
where $c_{\Lambda}(x,y)$ should be chosen to ensure the theory is finite  with the propagator vanishing for large momentum.
This has the advantage that the regularisation is built into the theory at the level of the action explicitly. Additionally, the regularised kinetic term in \eq{Reg_S} is the starting point for the non-perturbative renormalisation group in the Wilson-Polchinski formulation \cite{Polchinski:1983gv}. A standard choice for  $c_{\Lambda}(x,y)$ is the exponential cutoff
\beq
 c_{\Lambda}(x,y) = \frac{1}{(2 \pi)^d} \int d^d p \, e^{i p(x-y)} e^{- p^2/\Lambda^2} \,, 
\eeq
more generally we can set 
\beq
 c_{\Lambda}(x,y) = \frac{1}{(2 \pi)^d} \int d^d p \, e^{i p(x-y)} c(p^2/\Lambda^2)\,,
\eeq
for a monotonically decreasing function $c(p^2/\Lambda^2)$ that goes to zero sufficiently fast as its argument tends to infinity.

 \section{Perturbation theory}
 \label{Pert}
 In the previous section we worked in units $\hbar =1$. Putting back factors of $\hbar$ we can study the relation between genuine transformations \eq{hat_cov}  and effective transformations \eq{mean_cov} in perturbation theory.
If we evaluate $ \Gamma_{\hat{\phi}}[\phi] $ in the tree-level approximation starting from \eq{Gamma_2nd}  we have
\beq
 \Gamma_{\hat{\phi}}[\phi] =   S_{\hat{\phi}}[\phi]  + O(\hbar)\,,
\eeq
similarly 
\beq
 \Gamma_{\hat{\chi}}[\chi] = S[\chi]  + O(\hbar)\,.
\eeq
Then it follows from \eq{S_rel} that  \eq{GammaequalsGamma} holds at tree-level with
\beq
\chi[\phi] =  \hat{\chi}[\phi] + O(\hbar)\,.
\eeq
We can then continue to try to show that that a $\chi[\phi]$ exists such that \eq{GammaequalsGamma} holds order by order in $\hbar$.
This has been shown for  at one-loop order \cite{Cohen:2023ekv} where 
\beq \label{chi_one_loop}
\chi[\phi] = \hat{\chi}[\phi] + \hbar a[\phi]  + O(\hbar^2)\;,
\eeq 
with $a[\phi]$ being the one-loop correction. 
Let's determine the explicit form  $a[\phi]$ to see how this works.  First we have that
\beq\label{Gamma_one_loop}
\Gamma_\cc[\chi] = S[\chi] + \frac{\hbar}{2} \Tr \log S^{(2)}[\chi] + O(\hbar^2)\,,
\eeq
while
\beq \label{Gamma_zeta_one_loop}
\Gamma_\pp[\phi] = S_\pp[\phi] + \frac{\hbar}{2} \Tr \log S^{(2)}_\pp[\phi] + O(\hbar^2)\,,
\eeq 
where we use the superscript $(n)$ to denote the $n$th functional derivative of a functional with respect to its argument. 
The relation \eq{S_rel} can be written as 
\beq  \label{S_rel_2}
S_\pp[\hat{\phi}] = S[\hat{\chi}[\hat{\phi}]] - \hbar  \Tr \log \frac{\delta \hat{\chi}[\hat{\phi}]}{\delta \hat{\phi}}\,,
\eeq
which furthermore implies that the hessian $ S^{(2)}_\pp[\phi]$ is given by 
\beq
\frac{\delta^2 S_\pp[\phi]}{\delta \phi^x \delta \phi^y} =  \frac{\delta \hat{\chi}^z[\phi]}{\delta \phi^x}  S^{(2)}_{zw}[\hat{\chi}[\phi]]  \frac{\delta \hat{\chi}^w[\phi]}{\delta \phi^y}  +  \frac{\delta^2 \hat{\chi}^z[\phi]}{\delta \phi^x \delta \phi^y}  S^{(1)}_z[\hat{\chi}[\phi]]\, + O(\hbar)\; ,
\eeq
where here we adopt DeWitt notation such that a repeated coordinate index implies an integral over spacetime e.g. $\phi^x J_x = \int d^dx \phi(x) J(x)$.
Defining the ``connection'' by
\beq
\hat{C}^z_{xy}[\phi] = - \hat{Q}^{w}_x[\phi]     \frac{\delta^2 \hat{\chi}^z[\phi]}{\delta \phi^w \delta \phi^u}      \hat{Q}^{u}_y[\phi]\,,
\eeq
with
\beq \label{Qhat}
\hat{Q}^x_y[\hat{\phi}] \frac{\delta \hat{\chi}^y [\hat{\phi}]}{\delta \hat{\phi}^z} = \delta(x,z)\,,
\eeq
one sees that 
\beq
 \frac{\hbar}{2} \Tr \log S^{(2)}_\pp[\phi]  = \frac{\hbar}{2} \Tr \log  \left( S^{(2)}_{xy}[\hat{\chi}[\phi]]   - \hat{C}^z_{xy}[\phi]  S^{(1)}_z[\hat{\chi}[\phi]]\right) +  \hbar  \Tr \log \frac{\delta \hat{\chi}[\hat{\phi}]}{\delta \hat{\phi}}\,.
\eeq
Therefore using the above expression and \eq{S_rel_2} in \eqref{Gamma_zeta_one_loop} we get
\beq
\Gamma_\pp[\phi] = S[ \hat{\chi}[\phi]]  + \frac{\hbar}{2} \Tr \log  \left( S^{(2)}_{xy}[\hat{\chi}[\hat{\phi}]]   - \hat{C}^z_{xy}[\phi]  S^{(1)}_z[\hat{\chi}[\hat{\phi}]]\right)  + O(\hbar^2)\,,
\eeq 
on the other hand setting $\chi[\phi] = \hat{\chi}[\phi] + \hbar a[\phi]$ and using \eq{Gamma_one_loop}
\beq
\Gamma_\cc[\chi[\phi]] = S[ \hat{\chi}[\phi]]  +  \hbar S^{(1)}_x[\hat{\chi}[\phi]] a^x [\phi]+ \frac{\hbar}{2} \Tr \log S^{(2)}[ \hat{\chi}[\phi]] + O(\hbar^2)\,.
\eeq
Then the identity \eq{GammaequalsGamma} at one-loop holds if 
\beq
 S^{(1)}_x[\hat{\chi}[\phi]] a^x [\phi] =  \frac{\hbar}{2} \Tr \log  \left(\delta^x_y   -   D^{xu}[\hat{\chi}[\phi]]  \hat{C}^z_{uy}[\phi]  S^{(1)}_z[\hat{\chi}[\phi]]\right)\,,
\eeq
where $D^{xu}[\hat{\chi}]$ is the propagator i.e. the Greens function for  $S^{(2)}_{xy}$. Both sides of the above expression are proportional to $ S^{(1)}_x[\hat{\chi}[\phi]]$ so we can solve for $a^x$. Explicitly it is given by
\beq \label{a}
a^x = \frac{1}{2} \left[\log(1 + \hat{A})  \hat{A}^{-1}\right]^{y}\,_z \hat{B}^{zx}\,_y\,,
\eeq
with
\beq
\hat{A}^x\,_y = -   D^{xu}[\hat{\chi}[\phi]]  \hat{C}^z_{uy}[\phi]  S^{(1)}_z[\hat{\chi}[\hat{\phi}]]\,,
\eeq
and
\beq
\hat{B}^{xz}\,_y= -   D^{xu}[\hat{\chi}[\phi]]  \hat{C}^z_{uy}[\phi] \,,
\eeq
Hence at one-loop order the identity \eq{GammaequalsGamma} holds when we are given the change of variables $\hat{\chi}[\hat{\phi}]$. This demonstrates one direction of the correspondence at one-loop.

Next let us suppose we have an effective action $\Gamma[\chi]$ and we propose a change of variables of the mean field $\chi[\phi]$. Then we seek to find the change of variables of the fundamental fields $\hat{\chi}$ such that $\Gamma_\pp[\phi]= \Gamma_\cc[\chi[\phi]]$ generates the correlation functions of $\hat{\phi}$. Showing this demonstrates the correspondence \eq{GammaequalsGamma} in the other direction when we start from the effective change of variables.
At tree-level we have
\beq
\hat{\chi}[\hat{\phi}] = \chi[\hat{\phi}] + O(\hbar)\,.
\eeq  
At one-loop order we then write 
\beq \label{cc_one_loop}
\hat{\chi}[\hat{\phi}] = \chi[\hat{\phi}] + \hbar \hat{a}[\hat{\phi}]\,,
\eeq
So now we have again \eq{Gamma_one_loop} but $\Gamma_\pp[\phi]$ is given by
\beq\label{Gamma_one_loop_zeta}
\Gamma_\pp[\phi] = S[\chi[\phi]] + \frac{\hbar}{2} \Tr \log S^{(2)}[\chi[\phi]] + O(\hbar^2)\,,
\eeq
and what we want to show is that this is equal to a generating functional where the bare action is given by \eqref{S_rel_2} with some $\hat{\chi}[\hat{\phi}]$. At one loop order $S_{\hat{\phi}}[\hat{\phi}]$ is given by
\beq
S_{\hat{\phi}}[\hat{\phi}]= S[\chi[\pp]] +  \hbar S^{(1)}_x[\chi[\pp]] \hat{a}^x [\hat{\phi}]   - \hbar \Tr \frac{\delta \chi[\hat{\phi}]}{\delta \hat{\phi}} + O(\hbar^2) \;.
\eeq
Therefore we want to check that by inserting this expression in \eqref{Gamma_zeta_one_loop} we recover \eqref{Gamma_one_loop_zeta} for some $\hat{a}[\hat{\phi}]$ which we will determine. 
This time
\beq
\frac{\delta^2 S_\pp[\phi]}{\delta \phi^x \delta \phi^y} =  \frac{\delta \chi^z[\phi]}{\delta \phi^x}  S^{(2)}_{zw}[\chi[\phi]]  \frac{\delta \chi^w[\phi]}{\delta \phi^y}  +  \frac{\delta^2 \chi^z[\phi]}{\delta \phi^x \delta \phi^y}  S^{(1)}_z[\chi[\phi]] + O(\hbar)\,,
\eeq
thus we obtain 
\beq
 \frac{\hbar}{2} \Tr \log S^{(2)}_\pp[\phi]  = \frac{\hbar}{2} \Tr \log  \left( S^{(2)}_{xy}[\chi[\phi]]   - C^z_{xy}[\phi]  S^{(1)}_z[\chi[\phi]]\right) +  \hbar \frac{\delta \chi[\phi]}{\delta \phi}  + O(\hbar^2) 
\eeq
with $C^z_{xy}[\phi]$ defined similarly to $\hat{C}^z_{xy}[\phi]$ but with $\chi[\phi]$ in place of $\hat{\chi}[\phi]$ and therefore involving $Q^x_y[\phi]$ defined by
\beq \label{Q}
Q^x_y[\phi]  \frac{\delta  \chi^y[\phi] }{\delta \phi^z} = \delta(x,z)\,.
\eeq
Therefore, going this way, we have that $\hat{a}$ is determined by
\beq
 \frac{\hbar}{2} \Tr \log  \left( S^{(2)}_{xy}[\chi[\phi]]   - C^z_{xy}[\phi]  S^{(1)}_z[\chi[\phi]]\right)  +  \hbar S^{(1)}_x[\chi[\phi]] \hat{a}^x_\zeta [\hat{\phi}] =  \frac{\hbar}{2} \Tr \log S^{(2)}[\chi[\phi]] \,.
\eeq
So we have that
\beq \label{hata}
\hat{a}^x = - \frac{1}{2} \left[ \log(1 + A)  A^{-1}\right]^{y}\,_z B^{zx}\,_y \,,
\eeq
where $A$ and $B$ are defined similarly to $\hat{A}$ and $\hat{B}$ with $\chi[\phi]$ in place of $\hat{\chi}[\phi]$.
Thus, to one-loop order, we have shown the other direction of our correspondence  \eq{GammaequalsGamma}.
Note that \eq{a} and \eq{hata} can be derived from each other $a = -\hat{a}[\chi \to \hat{\chi}]$, which is obvious if we compare their definitions.

We could now continue to higher loop orders to show both the existence of $\chi[\phi]$ and $\hat{\chi}[\phi]$ given the other and to determine their forms order by order. Instead of doing so, we would like to give an implicit non-perturbative definition to  $\chi[\phi]$ and $\hat{\chi}[\phi]$ which imply their existence to all orders, without giving their explicit form. In this way we can be sure that higher loop corrections to \eq{chi_one_loop} and \eq{cc_one_loop} exist satisfying  \eq{GammaequalsGamma}.

\subsection*{One-loop example}
\label{one-loop-example}
Before moving beyond perturbation theory, it is illustrative to give an example at one-loop.
As an explicit example let's take the action to be given by 
\beq
S[\cc] =   \frac{1}{2} \int_{x,y} \partial_\mu \hat{\chi}(x) c^{-1}_{\Lambda}(x,y) \partial_\mu \hat{\chi}(y) +   \int_x V(\cc(x))\,,
\eeq
where $V$ is a potential and assume that $\hat{\chi}(x)$ only depends on $\pp$ and not its derivatives
\beq \label{Fhat}
\hat{\chi}(x) = \hat{F}(\pp(x))\,.
\eeq
In this case $a^x$ will be a function of $\pp$ and its derivatives. Here, for simplicity, we shall compute the terms that are only depend on $\phi$ such that
\beq
a^x = a^x_0(\phi(x)) + \dots \,,
\eeq
where the neglected terms are proportional to derivatives of $\phi$. The term $a^x_0(\phi(x)) $ is found by evaluating the general expression for $a$ at constant $\phi$.

Taking a functional derivative of \eq{Fhat} we find that
\beq
\frac{\delta \hat{\chi}^x [\hat{\phi}]}{\delta \hat{\phi}^y}  = \delta(x-y) \hat{F}'(\pp(x))\,,
\eeq
and hence
\beq
\hat{Q}(x,y) = \frac{1}{\hat{F}'(\pp(x))} \delta(x-y)\,,
\eeq
Taking a second derivative of $\cc$ we get 
\beq
\frac{\delta \hat{\chi}^x [\hat{\phi}]}{\delta \hat{\phi}^y  \hat{\phi}^z}  = \delta(x-y) \delta(x-z) \hat{F}''(\pp(x))\,,
\eeq
and so
\beq
\hat{C}^x_{yz}[\pp]  = -  \frac{\hat{F}''(\pp(x))}{\hat{F}'(\pp(x))^2} \delta(x-y) \delta(x-z) \,.
\eeq
Then the Hessian is given by
\beq
S^{(2)}_{xy} =  - \partial^2 c^{-1}(x,y) +  V''(\hat{F}(\pp)) \delta(x-y)\,,
\eeq
and we find that $\hat{A}[\phi]$ is given by
\beq
\hat{A}^x\,_y [\phi] = D^{xy}[\hat{F}(\phi)]  V'(F(\pp(x))) \frac{\hat{F}''(\pp(x))}{\hat{F}'(\pp(x))^2}\,.
\eeq
While
\beq
\hat{B}^{xz}\,_y=    D^{xz}[\hat{F}(\phi)] \frac{\hat{F}''(\pp(z))}{\hat{F}'(\pp(z))^2} \delta(z-y)\,.
\eeq
Putting everything together we find that\footnote{Here $\int_p = \int \frac{d^dp}{(2\pi)^d}$.}
\beq
 a^x_0(\phi) = \frac{1}{2} \int_p \log\left(1 +  \frac{c(p^2/\Lambda^2)}{p^2 + c(p^2/\Lambda^2) V''(\hat{F}(\phi))}  V'(\hat{F}(\phi)) \frac{\hat{F}''(\phi)}{\hat{F}'(\phi)^2} \right) \frac{1}{  V'(\hat{F}(\phi))}\,.
\eeq
Note that if $\hat{F}''(\phi)$ vanishes then $a^x_0$ is zero as it must be for a linear transformation. In the non-linear case the cutoff function $c(p^2/\Lambda^2)$ ensures the transformation is well defined.  Thus at one-loop and up to terms with derivatives of the field we obtain the effective change of variables
\beq \label{chi_ex_1loop}
\chi[\phi] =  \hat{F}(\phi) + \hbar   \frac{1}{2} \int_p \log\left(1 +  \frac{c(p^2/\Lambda^2)}{p^2 + c(p^2/\Lambda^2) V''(\hat{F}(\phi))}  V'(\hat{F}(\phi)) \frac{\hat{F}''(\phi)}{\hat{F}'(\phi)^2} \right) \frac{1}{  V'(\hat{F}(\phi))}\,.
\eeq

Now, if we instead start with an effective transformation $\chi[\phi] = F(\phi)$ then we can infer that the corresponding genuine transformation is
\beq
\cc[\pp] = \chi[\pp]-  \hbar \frac{1}{2} \int_p \log\left(1 +  \frac{c(p^2/\Lambda^2)}{p^2 + c(p^2/\Lambda^2) V''(F(\pp))}  V'(F(\pp)) \frac{F''(\pp)}{F'(\pp)^2} \right) \frac{1}{  V'(F(\pp))}\,.
\eeq
again up to derivatives of $\pp$ and $\hbar^2$ terms. If we send $\Lambda \to \infty$ in the above expression we get a divergent expression. On the other hand we can choose $F(\phi)$ to simplify the form of the effective action $\Gamma_{\pp}[\phi] = \Gamma_\chi[F(\phi)]$ which is free from divergencies.

\section{Implicit functions via initial value problems}
In the following two sections we shall define the changes of variables as solutions to initial value problems. 
Solving these initial value problems order by order in $\hbar$ we shall find linear initial value problems.
Here we review the solutions to such problems.

As a warm up let us consider a single function $s(\zeta)$ which satisfies linear differential equation
\beq
\frac{d  s(\zeta)}{d \zeta} = N(\zeta) s(\zeta) + \Xi(\zeta)\,, \,\,\,\,\,\,\,   s(\zeta_i) = s_i\,,
\eeq
where $N(\zeta)$ and $\Xi(\zeta)$ are two known functions. As long as $N(\zeta)$ is bounded then there is a unique solution to this class of initial value problems.
Indeed with
\beq
M(\zeta) = {\rm e}^{\int^\zeta_{\zeta_i}  d\zeta' N(\zeta') }\,,
\eeq
we have that the solution to the initial value problem is  
\beq
s(\zeta) = M(\zeta) \left( \int_{\zeta_i}^\zeta d\zeta' M^{-1}(\zeta) \Xi(\zeta) + s_i \right)\,.
\eeq

Now we consider the generalisation to a field variable $s^x(\zeta)$ defined by a linear initial value problem
\beq \label{ivp_phi}
\frac{d  s^x(\zeta)}{d \zeta} = N^x\,_y(\zeta) \phi^y(\zeta) + \Xi^x(\zeta)\,, \,\,\,\,\,\,\,   s^x(\zeta_i) = s^x_i\,,
\eeq
where $N^x\,_y(\zeta)$ is a known two-point function and  $\Xi^x(\zeta)$ is a known one-point function. 
Then we can define
\beq
M(\zeta) =\mathcal{T} {\rm e}^{\int^\zeta_{\zeta_i}  d\zeta' N(\zeta') }\,,
\eeq
where $\mathcal{T} $ is a $\zeta$-time ordering operator with later times always to the left of earlier times
and we understand $M(\zeta)$ and $N(\zeta)$ as continuous matrices.  
Then it follows that
\beq
\frac{d  M(\zeta)}{d \zeta} = N(\zeta)  M(\zeta)\,,
\eeq
or in component form
\beq
\frac{d  M^x\,_y(\zeta)}{d \zeta} = N^x\,_z(\zeta)  M^z\,_y(\zeta)\,.
\eeq
We then have that the solution to the initial value problem is  
\beq \label{ivp_sol}
s^x(\zeta) = M^{x}\,_y(\zeta) \left( \int_{\zeta_i}^\zeta d\zeta' (M^{-1})^y\,_z(\zeta) \Xi^z(\zeta) + s^y_i \right)\,.
\eeq
Although our changes of variables will be defined by non-linear initial value problems, at each order in $\hbar$ we will have a linear problem, and thus have a unique solution within perturbation theory.

 \section{Implicit effective field transformation}
In this section we assume the explicit form of the genuine transformation \eq{hat_cov} is known and aim to show that the result \eq{GammaequalsGamma} holds to all orders in perturbation theory by showing that there exists a corresponding  effective change of variables \eq{mean_cov}.  
This will show one direction of our proposed correspondence where we start from the genuine field transformation  and infer the existence of an effective transformation. To attack the problem we will work non-perturbatively and show that an implicit  effective change of variables can be defined as a solution to a non-linear initial value problem. While it is hard to know when a unique solution will exist to the non-linear problem, expanding in $\hbar$  we obtain a linear equation at each order $\hbar$.

We start by considering a family of field transformations $\hat{\chi}_\zeta[\pp]$ parameterised by a parameter $\zeta$. For an initial value $\zeta =\zeta_i$ we impose that 
\beq \label{pp_ic}
\hat{\chi}_{\zeta_i}[\pp] = \pp\,,
\eeq
 and for some final value $\zeta_f> \zeta_i$ we impose that $\hat{\phi}_\zeta[\cc]|_{\zeta =\zeta_f} =  \hat{\phi}[\cc]$. Denoting
\beq
\Gamma_\zeta[\phi] \equiv  \Gamma_{\hat{\phi}_\zeta}[\phi]\,,
\eeq
we have that  $\Gamma_\zeta[\phi]$ is defined via
\beq \label{Gamma_zeta_1}
e^{- \Gamma_\zeta[\phi] } = \int {\rm d} \hat{\chi}   \, e^{- S[\hat{\chi}]  + ( \hat{\phi}_\zeta[\hat{\chi}] - \phi) \cdot   \frac{ \delta \Gamma_\zeta[\phi]}{\delta \phi}      }\,.
\eeq
By construction $\zeta$ appears as a coupling for a one parameter family of effective actions such that the variation of $\zeta$ corresponds to a change of variables.  Thus $\zeta$ is by definition an inessential coupling \cite{Wegner:1974sla,Weinberg:1980gg,Baldazzi:2021ydj}: the value of $\zeta$ corresponds to a choice of which field variable we couple to the source $J$ and thus is purely conventional if ultimately we take $J=0$ and compute observables.

Our unknown, in verifying the correspondence,  is the effective change of variables $\phi_\zeta[\chi] $ or its inverse $\chi_\zeta[\phi]$. Here we are going to {\rm implicitly define}  $\phi_\zeta$ as a solution to an initial value problem
\beq \label{dotphi}
\frac{ \partial }{\partial \zeta} \phi_\zeta^x = \Psi^x_\zeta[\phi_\zeta] \,,
\eeq
where $\chi$ enters as the initial condition 
\beq
\phi_{\zeta_i} = \chi\,,
\eeq
and we will specify $\Psi^x_\zeta[\phi] $ shortly.
Defined this way, $\phi_\zeta[\chi]$ is the solution to the initial value problem which is unique as long as $\Psi^x_\zeta[\phi] $ is suitably continuous.
 Equally, using that $\phi_\zeta[\chi]$ and $\chi_\zeta[\phi]$ are inverse functions we have 
\beq \label{flow_chi}
\frac{\partial}{\partial \zeta} \chi_\zeta^x[\phi] = - \Psi_\zeta[\phi]  \cdot   \frac{\delta  }{\delta \phi }   \chi_\zeta^x[\phi] \,.
\eeq
In this case the equation \eq{flow_chi} is an initial value problem for $ \chi_\zeta^x[\phi]$ where by \eq{pp_ic} the initial condition is
\beq \label{chi_ic}
\chi_{\zeta_i}[\phi] = \phi\,.
\eeq
The form of $\Psi_\zeta[\phi]$ is determined  such that the solution to the initial value problem satisfies  
  \beq \label{Gammazeta}
\Gamma_\zeta[\phi] =  \Gamma[\chi_\zeta[\phi]]\,,
\eeq
and hence demonstrates the correspondence \eq{GammaequalsGamma}.
What we will show is that this specifies that
\beq\label{psi_1}
\Psi^x_\zeta[\phi]=  \left\langle \frac{\partial}{\partial \zeta}  \hat{\phi}^x_\zeta[\hat{\chi}] \right\rangle_{\pp_\zeta}[\phi] \,.
\eeq
To see that \eq{psi_1} implies \eq{Gammazeta} we note that  by differentiating \eq{Gamma_zeta_1}   with respect to $\zeta$ we find that $\Gamma_\zeta[\phi]$ itself solves an initial value problem given by (c.f. \eq{Gamma_deltapp})
 \beq \label{flow_Gamma}
\frac{\partial}{\partial \zeta}  \Gamma_\zeta[\phi] =  - \Psi_\zeta[\phi] \cdot \frac{\delta \Gamma_\zeta[\phi] }{\delta \phi}\,,
\eeq  
with the initial condition $\Gamma_{\zeta = 0}[\phi] = \Gamma_\cc[\phi]$ and $\Psi^x_\zeta[\phi]$ given by \eq{psi_1}.  Thus a functional   $\Gamma_\zeta[\phi]$ that satisfies  \eq{flow_Gamma},   with the initial condition provided by \eq{GammaSt}, will be equal to the effective action given by \eq{Gamma_zeta_1}, assuming uniqueness of the solution.
Then it is sufficient to show that \eq{Gammazeta}, with $\chi_\zeta[\phi]$ defined by its flow \eq{flow_chi},   obeys the flow equation  \eq{flow_Gamma}.
Inserting  \eq{Gammazeta}  into \eq{flow_Gamma} and using  \eq{flow_chi} we see that it does:
\bea
\frac{\partial}{\partial \zeta} \Gamma_\zeta[\phi]  &=&  \frac{\partial}{\partial \zeta}  \Gamma[\chi_\zeta[\phi]] \nn
&=&    \frac{\partial}{\partial \zeta} \chi_\zeta[\phi]          \cdot \frac{\delta \Gamma[\chi_\zeta] }{\delta \chi_\zeta }\nn
&=& - \Psi_\zeta[\phi]  \cdot   \frac{\delta  }{\delta \phi }   \chi_\zeta[\phi]    \cdot  \frac{\delta \Gamma[\chi_\zeta] }{\delta \chi_\zeta } \nn
& =&  - \Psi_\zeta[\phi] \cdot \frac{\delta \Gamma_\zeta[\phi] }{\delta \phi}\,.
\eea
Thus we have shown \eq{Gammazeta} and hence we confirm that \eq{GammaequalsGamma} holds non-perturbatively provided that unique solutions to the initial value problems exist. In other words, given the change of variables \eq{hat_cov} the effective change of variables \eq{mean_cov} is defined via the initial value problem \eq{dotphi} with \eq{psi_1}. 

To make the initial value problem explicit in terms of $\Gamma_{\zeta}$  let's define
\beq
 \hat{\Psi}^x_\zeta[\hat{\phi}_\zeta] = \frac{\partial}{\partial \zeta}\hat{\phi}^x_\zeta\,.
\eeq
which is known since we are starting from a given  $\hat{\phi}^x_\zeta$.
Then let us note that for any function of the field $\hat{F}[\pp]$ one can show that
\beq \label{Functional_identity}
\langle \hat{F}[\pp] \rangle_{\pp_\zeta}[\phi] =  \hat{F}\left[\phi + \frac{1}{\Gamma^{(2)}_\zeta[\phi]} \cdot \frac{\delta}{\delta \phi} \right]
\eeq
as we demonstrate in the appendix~\ref{App:1}.
In the rhs of  \eq{Functional_identity}, the operator acts on a factor of $1$ on the right with the derivative with respect to $\phi$ acting on the $\phi$ dependence of the operator itself. For example, if $\hat{F} = \hat{\phi}^x  \hat{\phi}^y $, then 
\bea \label{Functional_identity_example}
\langle  \hat{\phi}^x  \hat{\phi}^y \rangle_{\pp_\zeta}[\phi] &=& \left( \phi^x + \left(\frac{1}{\Gamma^{(2)}_\zeta[\phi]}\right)^{xz} \frac{\delta}{\delta \phi^z}\right)           \left( \phi^y + \left(\frac{1}{\Gamma^{(2)}_\zeta[\phi]}\right)^{yw} \frac{\delta}{\delta \phi^w}\right)\nn
&=&\left( \phi^x + \left(\frac{1}{\Gamma^{(2)}_\zeta[\phi]}\right)^{xz} \frac{\delta}{\delta \phi^z}\right)          \phi^y \nn
&=& \phi^x \phi^y  + \left(\frac{1}{\Gamma^{(2)}_\zeta[\phi]}\right)^{xy}  \,.       
\eea
Then we can write
\beq \label{Psi_hat_Psi}
\Psi^x_\zeta[\phi] = \hat{\Psi}^x_\zeta\left[\phi + \frac{1}{\Gamma^{(2)}_\zeta[\phi]} \cdot \frac{\delta}{\delta \phi} \right]\,.
\eeq
This expression gives a multi-loop expression in terms of functional derivatives of the effective action $\Gamma_\zeta$ and of $\hat{\Psi}^x$ with the diagrams involving full 1PI vertices and propagators. Expanding in loops we obtain 
\beq \label{Psi_loop}
\Psi^x_\zeta[\phi]  = \hat{\Psi}^x_\zeta\left[\phi + \frac{1}{\Gamma^{(2)}_\zeta[\phi]} \cdot \frac{\delta}{\delta \phi} \right]  = \hat{\Psi}^x_\zeta\left[\phi \right] + \frac{1}{2} \left( \frac{1}{\Gamma^{(2)}_\zeta[\phi]} \right)^{yz}  \frac{\delta^2  \hat{\Psi}^x[\phi]}{\delta \phi^y \delta \phi^z}   + {\rm higher\,loops}\,.
\eeq
(which can be further expanded in $\hbar$ by expanding $\Gamma_\zeta^{(2)}$).
Going to higher loop orders we will encounter all $n$-point functions of  $\Gamma_\zeta$ and $\hat{\Psi}^x$.

Now let's write
\beq
\phi_\zeta= \pp_\zeta + \sum_{\ell=1}^{\infty} \hbar^\ell \phi_{\ell,\zeta}\,,
\eeq
and 
\beq
\Psi^x_\zeta[\phi] = \hat{\Psi}^x_\zeta[\phi] + \sum_{\ell=1}^{\infty} \hbar^\ell \Psi_{\ell, \zeta}[\phi]\,,
\eeq
and insert these into \eq{dotphi}. Expanding in $\hbar$ and comparing coefficients of $\hbar^\ell$ we will get a set of equations  which are linear in $ \phi_{\ell, \zeta}$
\beq
\frac{\partial}{\partial \zeta} \phi_{\zeta, \ell}   =\left. \frac{\delta \hat{\Psi}^x_\zeta}{\delta \hat{\phi}^y} \right|_{\pp = \pp_\zeta} \phi^y_{\zeta, \ell} +  \Xi_\ell^x(\zeta)\,,
\eeq
where $\Xi_\ell^x(\zeta)$ depends on $\phi_{\ell'}$ for $\ell'<\ell$ and $\pp_\zeta$. Thus starting at order $\ell =1$ we can solve these equations iteratively  and thus only encounter linear initial value problems of the form \eq{ivp_phi} where at each order $s(\zeta) = \phi_{\ell,\zeta}$.   We can integrate the initial value problems to obtain the solution of the form \eq{ivp_sol}. 
Thus we conclude that explicit expressions exist for each  $\phi_{\ell,\zeta}$.

\subsection*{One-loop example revisited}

Let us now return to our one-loop example \ref{one-loop-example} to check that our non-perturbative implicit change of variables agrees with our explicit one-loop calculation by checking that \eq{chi_ex_1loop} satisfies the flow \eq{flow_chi}. There we took the change of variables to be for the form $\hat{\chi}^x[\pp] = \hat{F}(\pp(x))$ and we now let  $\hat{F}(\pp(x))$ depend on $\zeta$ such that 
\beq
\hat{\chi}^x_\zeta[\pp] = \hat{F}_\zeta(\pp(x))\,.
\eeq
It follows that 
\beq
\frac{\partial}{\partial \zeta} \hat{\phi}^x_\zeta[\cc] |_{\cc = \cc_\zeta[\phi]}=     \hat{\Psi}^x_\zeta[\hat{\phi}] = \hat{f}_\zeta( \hat{\phi}(x))\,,
\eeq
with 
\beq
\hat{f}( \hat{\phi}(x)) = - \frac{ \frac{\partial}{\partial \zeta} \hat{F}_\zeta(\pp(x))}{   \hat{F}_\zeta'(\pp(x))}\,.
\eeq
Then using \eq{Psi_loop} one finds that to one-loop order 
\beq \label{Psi_1_loop}
\Psi^x_\zeta[\phi] = \hat{f}_\zeta( \phi(x)) + \frac{\hbar}{2} D^{xx}_{\zeta}[\phi] \hat{f}''( \phi(x)) \,,
\eeq
where $D^{xy}_{\zeta}$ is the tree-level propagator for the action $S_\zeta[\pp] \equiv S_{\hat{\phi}_\zeta}[\pp] =  S[\hat{\chi}_\zeta[\hat{\phi}]] $.
Up to terms involving derivatives of the field
\beq \label{Dxx_no_ders}
D^{xx}_{\zeta}[\phi]  = \int_p \frac{1}{  \left[ p^2 c^{-1}(p^2/\Lambda^2) +  V''(\hat{F}_\zeta(\pp(x))) \right]  \hat{F}_\zeta'(\pp(x))^2  +  V'(\hat{F}_\zeta(\pp(x)))  \hat{F}_\zeta''(\pp(x)) }\,,
\eeq
One can then check that \eq{chi_ex_1loop}, with $\hat{F}= \hat{F}_\zeta$  satisfies the flow equation \eq{flow_chi} with \eq{Psi_1_loop} and \eq{Dxx_no_ders}. Indeed, let's define from  \eq{chi_ex_1loop}
\beq
F_\zeta(\phi) =   \hat{F}_\zeta(\phi) + \hbar   \frac{1}{2} \int_p \log\left(1 +  \frac{c(p^2/\Lambda^2)}{p^2 + c(p^2/\Lambda^2) V''(\hat{F}_\zeta(\phi))}  V'(\hat{F}_\zeta(\phi)) \frac{\hat{F}_\zeta''(\phi)}{\hat{F}_\zeta'(\phi)^2} \right) \frac{1}{  V'(\hat{F}_\zeta(\phi))}\,,
\eeq
which is our approximation of the effective change of variables $\chi_\zeta[\phi]$.
Then we should have, from \eq{flow_chi} and \eq{Psi_1_loop}, that
\beq
\hat{f}_\zeta( \phi(x)) + \frac{\hbar}{2} D^{xx}_{\zeta}[\phi] \hat{f}''_\zeta( \phi(x))  = - \frac{ \partial_\zeta F_\zeta(\pp(x))}{   F_\zeta'(\pp(x))} + O(\hbar^2, \partial \phi)\,.
\eeq
Using the forms of $\hat{f}_\zeta( \hat{\phi}(x)) $ and $F_\zeta(\phi) $ in terms of $\hat{F}_\zeta(\phi)$ and its $\phi$ and $\zeta$ derivatives it is straight forward to check that the identity holds. 

\section{Implicit genuine field transformation }

As in the last section we again consider a one parameter set  of effective actions $\Gamma_\zeta[\phi]$. 
Here we assume that the effective actions $\Gamma_\zeta[\phi]$ are related to $\Gamma_{\hat{\chi}}[\chi]$ by a known change of variables $\chi \to \chi_{\zeta}[\phi]$ such that \eq{Gammazeta} holds and we assume $S[\cc]$ is known. So in this case we know $\Psi_\zeta[\phi]$ as it is given by \eq{dotphi} and \eq{flow_Gamma} holds.
What we want to show is that we can define implicitly actions $\hat{S}_\zeta[\pp]$ such that 
\beq \label{Gamma_zeta_2}
e^{- \Gamma_\zeta[\phi] } = \int d \hat{\phi}   \, e^{-S_\zeta[\hat{\phi} ]   + ( \hat{\phi} - \phi) \cdot   \frac{ \delta \Gamma_\zeta[\phi]}{\delta \phi}      }\,.
\eeq 
Since $S_\zeta[\pp]$ for different values of $\zeta$ must be related by a change of variables we must have
\beq\label{S_hat_flow}
\frac{\partial}{\partial \zeta} e^{-S_\zeta[\hat{\phi} ]} = - \frac{\delta}{\delta \pp^x}  \left( \hat{\Psi}_\zeta^x[\pp]  e^{-S_\zeta[\hat{\phi} ]} \right) \,,
\eeq
with the initial condition 
\beq
S_{\zeta_i}[\hat{\phi} ] = S[\pp]\,,
\eeq
for some $\hat{\Psi}_\zeta^x[\pp]$ that we wish to determine. The solution to the initial value problem
 \beq \label{flow_pp}
\frac{\partial}{\partial \zeta} \pp_\zeta^x =  \hat{\Psi}_\zeta[\pp]  \,,
\eeq
 with
 \beq
\pp_{\zeta_i}^x[\cc] = \cc\,,
\eeq
 then defines the change of variables $\pp_\zeta[\cc]$. 
What we seek is a closed form expression for  $ \hat{\Psi}_\zeta[\pp]$ in terms of the knowns i.e. in terms of $S[\cc]$ and $\Psi_\zeta[\phi]$.
To this end let's now introduce the generator of connected correlation functions by the Legendre transform
\beq
W_\zeta[J] = \sup_{\phi} ( - \Gamma_\zeta[\phi] + J\cdot\phi)\,,
\eeq 
which implies that  
\beq
\phi^x = \frac{\delta W_\zeta[J] }{\delta J_x}\,,    \,\,\,\,\,\,\,     \frac{\delta \Gamma_\zeta[\phi] }{\delta \phi^x} = J\,,
\eeq
such that
\beq \label{WSzeta}
e^{W_\zeta[J]} = \int d\pp e^{-S_\zeta[\pp]+ J \cdot \pp} \,.
\eeq
By \eq{flow_Gamma} we obtain the flow of $W_\zeta[J]$
\beq \label{flow_W}
\frac{\partial}{\partial \zeta} W_\zeta[J] =\tilde{\Psi}_\zeta[J]  \cdot J\,,
\eeq
 where
\beq
\tilde{\Psi}_\zeta[J]  =  \Psi_{\zeta}\left[ \frac{\delta W_\zeta}{\delta J}\right] \,.
\eeq
Then the $W_\zeta[J]$ which satisfies this equation with  the initial condition at $\zeta =\zeta_i$ will generate the same correlation functions of $\Gamma_\zeta$.
Our aim is to find the form of  $\hat{\Psi}_\zeta[\pp]$ such that \eq{WSzeta} obeys \eq{flow_W} given  \eq{S_hat_flow}.
As we will see this form is given by 
\beq \label{Psi_hat}
\hat{\Psi}_\zeta[\pp]  = e^{S_\zeta[\pp]}  \tilde{\Psi}_{\zeta}\left[- \frac{\delta}{\delta \hat{\phi}} \right] e^{-S_\zeta[\pp]}\,,  
\eeq
which
can be written as
\beq \label{Psi_hat2}
\hat{\Psi}_\zeta[\pp] = \tilde{\Psi}_\zeta\left[S_\zeta^{(1)}[\pp]   -  \frac{\delta}{\delta\hat {\phi}} \right]\,,
\eeq
as is shown in the appendix~\ref{App:1}.
Then from \eq{WSzeta} we have
\beq 
\frac{\partial}{\partial \zeta} W_\zeta[J]  =   e^{-W_\zeta[J]} \int d\pp \frac{\partial}{\partial \zeta}  e^{-S_\zeta[\pp]+ J \cdot \pp} \,.
\eeq
Using \eq{S_hat_flow}  and integrating by parts we get 
\bea
\frac{\partial}{\partial \zeta} W_\zeta[J]  &=&  - e^{-W_\zeta[J]} \int d\pp       \frac{\delta}{\delta \pp^x}  \left( \hat{\Psi}_\zeta^x[\pp]  e^{-S_\zeta[\hat{\phi} ]} \right)   e^{J \cdot \pp}\nn
&=& e^{-W_\zeta[J]} \int d\pp  \,    J_x  \left( \hat{\Psi}_\zeta^x[\pp]  e^{-S_\zeta[\hat{\phi} ]} \right)   e^{J \cdot \pp}\,.
\eea
Then we use \eq{Psi_hat} 
\bea
\frac{\partial}{\partial \zeta} W_\zeta[J]  &=&  e^{-W_\zeta[J]} \int d\pp      J_x  \left( \tilde{\Psi}^x_{\zeta}\left[- \frac{\delta}{\delta \hat{\phi}} \right] e^{-S_\zeta[\pp]}\right)   e^{J \cdot \pp}\nn
&=&  e^{-W_\zeta[J]} \int d\pp      J_x  \left( \tilde{\Psi}^x_{\zeta}\left[J \right] e^{-S_\zeta[\pp]}\right)   e^{J \cdot \pp}\nn
&=& J_x  \tilde{\Psi}^x_{\zeta}\left[J \right]\,,
\eea
which is what we required. 
So we can find $S_\zeta[\pp]$ by solving the initial value problem 
\beq\label{S_hat_flow_explicit}
\frac{\partial}{\partial \zeta} e^{-S_\zeta[\hat{\phi} ]} = - \frac{\delta}{\delta \pp^x}  \left( e^{-S_\zeta[\hat{\phi} ]}  \tilde{\Psi}_\zeta^x\left[S_\zeta^{(1)}[\pp]   -  \frac{\delta}{\delta\hat {\phi}} \right] \right) \,.
\eeq
We conclude that we have non-linear initial value problem for $S_{\zeta}[\pp]$ which defines it non-perturbatively if a unique solution exists.
Putting the solution of this equation into the flow for $\pp_\zeta[\cc]$ given by \eq{flow_pp} we can solve for  $\pp_\zeta[\cc]$ as well.
As in the last section, if we expand in $\hbar$ we get linear initial value problems at each order and hence at each order we will get a unique solution. Thus again we confirm \eq{GammaequalsGamma}.

At one-loop order we find that 
\beq
\hat{\Psi}_\zeta[\pp]  =  \tilde{\Psi}^x[S_\zeta^{(1)}[\pp]]   + \hbar \frac{1}{2}  S_{\zeta ,yz}[\pp]   D^{yz w}_\zeta[\pp] \Psi^{x}_{\zeta ,w}[\pp]  -  \hbar\frac{1}{2}    \Psi^x_{\zeta,yv}[\pp]  D^{yv}_{\zeta}[\pp]   + O(\hbar^2)\,,
\eeq 
where
\beq
D^{yz w}_\zeta[\pp]  = D^{yx}_\zeta D^{zv}_\zeta D^{w u}_\zeta S_{\zeta, xvu}\,,
\eeq
using the expression for the one loop effective action $\Gamma_\zeta[\phi]$ and that
\beq
  \tilde{\Psi}^x[\Gamma_\zeta^{(1)}[\pp]] =  \Psi^x[\pp] \,,
\eeq
 this reduces to 
\beq
\hat{\Psi}_\zeta[\pp]  =  \Psi^x[\pp]    - \frac{\hbar}{2}    \Psi^x_{\zeta,yz}[\pp]  D^{yz}_{\zeta}[\pp]    + O(\hbar^2)\,.
\eeq 
The loop is regularised by our regularisation of the propagator. We observe that this equation is consistent with  \eq{Psi_loop}.

\section{The equivalence theorem }

In this section we will show that amplitudes $\mathcal{A}(\bar{p}_1, \dots \bar{p}_n)$ are independent of the choice of variables subject to locality assumptions. This demonstrates that the equivalence theorem holds,  provided the change of variables is suitably local, without using perturbation theory. 
 The amplitudes are defined for a theory with a single massive pole in the propagator as the momentum go on shell
\beq \label{momentum_shell}
 \bar{p}^2_i = m^2\,.
\eeq
By the LSZ reduction formula  $\mathcal{A}(\bar{p}_1, \dots \bar{p}_n)$ is the residue of the poles off the renormalised correlation functions.   
Assuming \eq{GammaequalsGamma},   \cite{Cohen:2022uuw,Cohen:2023ekv} showed that the amplitudes are independent of the choice of variables using a recursive formula for the amplitudes. Since we have shown   \eq{GammaequalsGamma} this  gives a demonstration of the equivalence theorem.  
Here we will give a complementary demonstration of this equivalence by showing that amplitudes are independent of the value of the inessential couplings: 
\beq \label{Equivalence}
\frac{\partial}{\partial \zeta} \mathcal{A}(\bar{p}_1, \dots \bar{p}_n) =0\,.
\eeq
For this we will not need to use \eq{GammaequalsGamma} directly but only \eq{flow_Gamma} from which the former result was also obtained. Since  \eq{flow_Gamma} holds non-perturbatively so does \eq{Equivalence}.

The amplitudes   $\mathcal{A}(\bar{p}_1, \dots \bar{p}_n)$ are, up to a normalisation factor, the residues of the poles of connected momentum space correlation functions when the momentum are put on-shell. In order to be sure that \eq{Equivalence} holds we will need to make an assumption about the locality of $\Psi^x_\zeta[\phi]$ namely we will assume the pole free condition
\beq \label{pole_free}
 \int_x (p^2 - m^2) e^{ix p} \Psi^x_\zeta[\phi] \to 0  \,,  \,\,\,\,\,\,{\rm  as} \,\,\,\,  p^2 \to \bar{p}^2 = m^2 
\eeq
This is a weak locality condition that is satisfied if $\Psi^x_\zeta[\phi]$ is strictly local while allowing for some weak non-localities. To give an example of where the condition fails we suppose
 \beq
 \Psi^x_\zeta[\phi] \stackrel{!}{=}   -\frac{1}{\Box_x + m^2} \phi^2(x) 
\eeq
where $\Box = \partial_t^2 - \nabla^2$ is the d'Alembertian.
Then
\beq
 \int_x (\bar{p}^2 - m^2) e^{ix \bar{p}} \Psi^x_\zeta[\phi]    \stackrel{!}{=}   \int_x e^{ix \bar{p}}    \phi^2(x)  \neq 0\,.
\eeq
In this case the non-local form of $\Psi^x_\zeta[\phi]$ will invalidate our proof.
On the other hand there are non-local forms  $\Psi^x_\zeta[\phi]$ which satisfy \eq{pole_free} and thus the transformations can be non-local.
For a recent discussion on non-local field redefinitions see \cite{Cohen:2024fak}.

To begin, first we study how the on-shell propagator depends on $\zeta$. 
The inverse propagator is given by
%\footnote{Similarly to the action we keep factors of $i$ inside $Z_\zeta(\Box)$ i.e.  $Z_\zeta(-\Box) = -i Z^L_\zeta(-\Box)$ where $Z^L_\zeta(-\Box)$ is the standard Lorentzian wave-function renormalisation.  }
\beq \label{inverse_propagator}
\Gamma^{(2)}_\zeta[\bar{\phi}_\zeta] = i Z_\zeta(-\Box)( \Box + m^2)
\eeq
where $ Z_\zeta(p^2)$ is the momentum dependent wave function renormalisation. Since we are at vanishing source $\bar{\phi}_\zeta$ satisfies the equation of motion
\beq
\Gamma^{(1)}_\zeta[\bar{\phi}_\zeta] = 0\,
\eeq
and is assumed to be constant over spacetime, hence the simple form of \eq{inverse_propagator}.
The propagator is the Greens function of \eq{inverse_propagator} which in momentum space has a pole with a residue
\beq
R_\zeta = \frac{1}{Z_\zeta(m^2)}\,.
\eeq
The renormalised field is then $R_\zeta^{-1/2}\pp_\zeta[\cc]$.
For later use we define $G^{xy}$  as the {\it normalised} propagator
\beq \label{G}
-( \Box_x + m^2)G^{xy} = \delta(x,y)
\eeq 
i.e. without the factor of $1/Z_\zeta(-\Box)$ (this simplifies some steps since $G^{xy}$ is independent of $\zeta$). Next we wish to relate the flow of $Z_\zeta$ to $\Psi^z_\zeta[\phi]$.  First we note that
\beq
\bar{\phi}_\zeta = \phi_\zeta[\bar{\chi}] 
\eeq
where 
\beq
\Gamma^{(1)}_0[\bar{\chi}] = 0\,
\eeq
 By taking one derivative of  \eq{flow_Gamma} with respect to $\phi(x)$  and setting $\phi = \bar{\phi}_\zeta$ one finds that 
\beq \label{flowbarphi}
\frac{\partial}{\partial \zeta}  \bar{\phi}_\zeta  = \Psi_\zeta[\bar{\phi}_\zeta]
\eeq
Next by differentiating \eq{flow_Gamma} twice with respect to $\phi$ and setting $\phi = \bar{\phi}_\zeta$ we get
\beq
\frac{\partial}{\partial \zeta}  \frac{\delta^2 \Gamma_\zeta[\bar{\phi}_\zeta]}{\delta \phi^x \delta \phi^y}  =    - \Psi^z_\zeta[\bar{\phi}_\zeta] \cdot \frac{\delta \Gamma_\zeta[\bar{\phi}_\zeta] }{\delta \phi^z \delta \phi^x \delta\phi^y} -  \frac{\delta \Psi^z_\zeta}{\delta \phi^x}[\bar{\phi}_\zeta] \cdot \frac{\delta \Gamma_\zeta[\bar{\phi}_\zeta] }{\delta \phi^z \delta\phi^y} - \frac{\delta \Psi^z_\zeta}{\delta \phi^y}[\bar{\phi}_\zeta] \cdot \frac{\delta \Gamma_\zeta[\bar{\phi}_\zeta] }{\delta \phi^z \delta\phi^x}\,.
\eeq
By \eq{flowbarphi} the first term on the rhs of the above expression can be brought to the lhs to form a total derivative
\beq \label{flow_inverse_propagator}
\frac{d}{d \zeta}  \frac{\delta^2 \Gamma_\zeta[\bar{\phi}_\zeta]}{\delta \phi^x \delta \phi^y} =  -  \frac{\delta \Psi^z_\zeta}{\delta \phi^x}[\bar{\phi}_\zeta] \cdot \frac{\delta \Gamma_\zeta[\bar{\phi}_\zeta] }{\delta \phi^z \delta\phi^y} - \frac{\delta \Psi^z_\zeta}{\delta \phi^y}[\bar{\phi}_\zeta] \cdot \frac{\delta \Gamma_\zeta[\bar{\phi}_\zeta] }{\delta \phi^z \delta\phi^x}
\eeq
defining $\eta_\zeta(-\Box)$ by 
\beq \label{eta_zeta}
\frac{\delta \Psi^z_\zeta}{\delta \phi^y}[\bar{\phi}_\zeta] =\frac{1}{2} \eta_\zeta(-\Box) \delta(x,y)
\eeq
 and inserting \eq{inverse_propagator} into \eq{flow_inverse_propagator} we can solve for the flow of $Z_\zeta(-\Box)$.
 This gives 
\beq
\frac{\partial}{\partial \zeta} Z_\zeta(-\Box) = - Z_\zeta(-\Box) \eta(-\Box)\,.
\eeq

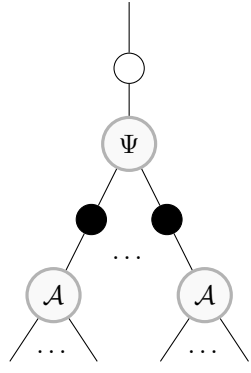
\begin{figure}
\begin{tikzpicture}
[bub/.style={circle, draw=gray!60, fill=gray!5, very thick, minimum size=7mm},prop/.style={circle, draw=black!100, fill=black!100, minimum size=4mm},inv/.style={circle, draw=black!100, fill=white, minimum size=4mm}]
%Nodes
\node[inv] (Inv) at     (1,3) {};
\node[bub] (Psi)  at    (1,2)     {$\Psi$};
\node[bub]   (X)  at    (0,0)     {$\mathcal{A}$};
\node[bub] (Y)  at    (2,0)     {$\mathcal{A}$};
\node[] (DOTS)  at    (1,0.5)   {$\dots$};
\node[prop] (Z)  at    (0.5,1)    {};
\node[prop] (W)  at    (1.5,1)   {};
\node[] (U0)  at    (1,4)   {};
\node[] (U1)  at    (-0.66,-1)   {};
\node[] (U2)  at    (-0.33,-1)   {};
\node[] (U3)  at    (0,-1)   {};
\node[] (U4)  at    (0.33,-1)   {};
\node[] (U5)  at    (0.66,-1)   {};
\node[] (V1)  at    (2.66,-1)   {};
\node[] (V2)  at    (2.33,-1)   {};
\node[] (V3)  at    (2,-1)   {};
\node[] (V5)  at    (1.33,-1)   {};
\node[] (V4)  at    (1.66,-1)   {};
\node[] (DOTS2)  at    (2,-0.75)   {$\dots$};
\node[] (DOTS1)  at    (0,-0.75)   {$\dots$};
% Lines
\draw[-]  (X) --(Z);
\draw[-]  (Y) --(W);
\draw[-]  (Z) --(Psi);
\draw[-]  (Psi) --(Inv);
\draw[-]  (Inv) --(U0);
\draw[-]  (W) --(Psi);
\draw[-]  (U1) --(X);
%\draw[-]  (U2) --(X);
%\draw[-]  (U3) --(X);
%\draw[-]  (U4) --(X);
\draw[-]  (U5) --(X);
\draw[-]  (V1) --(Y);
%\draw[-]  (V2) --(Y);
%\draw[-]  (V3) --(Y);
%\draw[-]  (V4) --(Y);
\draw[-]  (V5) --(Y);
\end{tikzpicture}
\caption{ \label{General_Diagram}The momentum space diagrammatic representation of each contribution to the lhs of \eq{Identity}, which are written in \eq{Terms}. The white circle is the inverse propagator $p^2_j -m^2$  which connects to the index of $\Psi$. The functional derivatives of $\Psi$ are represented  by the accordingly labeled vertex with each lower leg denoting a derivative. The black circles are propagators $1/(q^2 - m^2)$ (c.f. \eq{G}). The functional derivatives of $\mathcal{A}$ are represented by the vertices labeled accordingly.
The dots indicate the different varieties of such diagrams with different numbers of $\mathcal{A}$ vertices each with a different number of legs.   }
\end{figure}

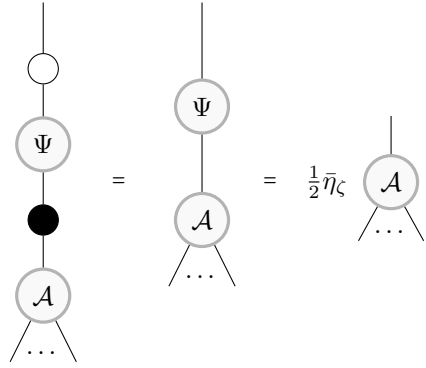
\begin{figure}

\begin{tikzpicture}
[bub/.style={circle, draw=gray!60, fill=gray!5, very thick, minimum size=7mm},prop/.style={circle, draw=black!100, fill=black!100, minimum size=4mm},inv/.style={circle, draw=black!100, fill=white, minimum size=4mm}]
%Nodes
\node[inv] (Inv) at     (1,3) {};
\node[bub] (Psi)  at    (1,2)     {$\Psi$};
\node[prop] (Z)  at    (1,1)    {};
\node[bub]   (X)  at    (1,0)     {$\mathcal{A}$};
\node[] (U0)  at    (1,4)   {};
\node[] (U1)  at    (1.5,-1)   {};
\node[] (U2)  at    (0.5,-1)   {};
\node[] (DOTS2)  at    (1,-0.75)   {$\dots$};

\node[]  at    (2,1.5)     {$=$};

%\node[inv] (Inv2) at     (3.1,3) {};
\node[] (U02)  at    (3.1,4)   {};
\node[bub] (Psi2)  at    (3.1,2.5)     {$\Psi$};
%\node[prop] (Z2)  at    (3.1,1)    {};
\node[bub]   (X2)  at    (3.1,1)     {$\mathcal{A}$};
\node[] (U12)  at    (3.6,0)   {};
\node[] (U22)  at    (2.6,0)   {};
\node[] (DOTS22)  at    (3.1,0.25)   {$\dots$};

\node[]  at    (4,1.5)     {$=$};

\node[]  at    (4.75,1.5)     {$\frac{1}{2} \bar{\eta}_\zeta$};

\node[] (U03)  at    (5.6,2.5)   {};
\node[bub]   (X3)  at    (5.6,1.5)     {$\mathcal{A}$};
\node[] (U13)  at    (5.1,0.6)   {};
\node[] (U23)  at    (6.1,0.6)   {};
\node[] (DOTS22)  at    (5.6,0.85)   {$\dots$};

%\node[] (DOTS1)  at    (0,-0.75)   {$\dots$};
% Lines
\draw[-]  (X) --(Z);
\draw[-]  (Z) --(Psi);
\draw[-]  (Psi) --(Inv);
\draw[-]  (Inv) --(U0);
%\draw[-]  (W) --(Psi);
\draw[-]  (U12) --(X2);
\draw[-]  (U22) --(X2);

\draw[-]   (U02)--(Psi2);
\draw[-]   (Psi2)--(X2);
\draw[-]  (U1) --(X);
\draw[-]  (U2) --(X);

\draw[-]   (U03)--(X3);
\draw[-]   (U13)--(X3);
\draw[-]   (U23)--(X3);

\end{tikzpicture}
\caption{ \label{non-zero-diagram} The non-zero contributions to the lhs of \eq{Identity} is expressed diagrammatically and sum to give \eq{non-zero}. The propagator and its inverse cancel since they both are have the same momentum running through them leading to the simpler second expression (c.f. \eq{GBox}).
Taking the momentum on-shell gives the final expression. These diagrams add up to give exactly the rhs of \eq{Identity}.  }
\end{figure}

Now we turn to the higher $n$-point functions. The renormalised connected correlation functions are given by
\beq
\langle R_\zeta^{-1/2} \pp^{x_1}_\zeta[\cc] \dots  R_\zeta^{-1/2}\pp^{x_n}_\zeta[\cc] \rangle_c =  \left. R_\zeta^{-n/2} \frac{\delta W_\zeta}{\delta J_{x_1} \dots \delta J_{x_n}}\right|_{J=0} \,.   
\eeq
 When expressed in terms of the one-particle irreducible diagrams, the renormalised correlation functions have legs consisting for a factor of the propagator  $G^{xy}$. 
Fourier transforming the connected diagrams and taking the external momentum on shell $\bar{p}_i^2 = m^2$  we will get poles for each external momentum. The residue of this pole is the amplitude.
Equivalently, removing the factors of $G^{xy}$ from a connected correlation function we get the amputated diagram $\A_{\zeta}(x_1,\dots x_n)$ whose Fourier transform is the amplitude.  Thus it is most straightforward to work with a generating functional of amputated correlation functions
\beq \label{A}
\mathcal{A}_\zeta[\psi]  = W_\zeta[  (-\Box -m^2) \psi] \,,
\eeq
Then by the reduction formula we have
\beq
\mathcal{A}(\bar{p}_1, \dots \bar{p}_n) =  R^{-n/2}_\zeta \left.\left( \prod_{i=1}^n \int_{x_i}   e^{{\rm i} \bar{p}_i x_i} \right)   \frac{ \delta^n \mathcal{A}_\zeta[\psi] }{\delta \psi^{x_1} \dots \delta \psi^{x_n}} \right|_{\psi = 0}\,,
\eeq
where the momentum $\bar{p}_i$ satisfy \eq{momentum_shell}.
Now we note that by \eq{A} and \eq{flow_W} we have
\beq
\frac{\partial}{\partial \zeta} \mathcal{A}_\zeta[\psi] = \Psi_{\zeta }\left[ G \cdot \frac{\delta \mathcal{A}}{\delta \psi}\right]  \cdot ( -\Box - m^2) \psi\,.
\eeq
Defining 
\beq
\bar{\eta}_\zeta =  \frac{1}{R_\zeta} \frac{\partial}{\partial \zeta} R_\zeta = \eta_\zeta(m^2)\,,
\eeq
we have that
\bea
\frac{\partial }{\partial \zeta}   \mathcal{A}(\bar{p}_1, \dots \bar{p}_n)  &=& - \frac{n\bar{\eta}_\zeta}{2}  \mathcal{A}(\bar{p}_1, \dots \bar{p}_n) \nn
&&+  R^{-n/2}_\zeta \left. \left( \prod_{i=1}^n \int_{x_i}   e^{{\rm i} p_i x_i}  \right)  \frac{ \delta^n  }{\delta \psi^{x_1} \dots \psi^{x_n}}  \left(    \Psi_{\zeta }^x\left[ G \frac{\delta \mathcal{A}}{\delta \psi}\right]   ( -\Box_x - m^2) \psi^x   \right)  \right|_{\psi = 0,p_i\to \bar{p}_i }\,.
\eea
Hence to show the equivalence theorem in the differential form \eq{Equivalence} we have to demonstrate that 
\beq \label{Identity}
R^{-\frac{n}{2}}_\zeta \left.  \left(  \prod_{i=1}^n \int_{x_i}   e^{{\rm i} p_i x_i} \right)   \frac{ \delta^n  }{\delta \psi^{x_1} \dots \psi^{x_n}}  \left(    \Psi_{\zeta }^x\left[ G \frac{\delta \mathcal{A}}{\delta \psi}\right]   ( -\Box_x - m^2) \psi^x   \right)  \right|_{\psi = 0,p_i\to \bar{p}_i} =   \frac{n}{2} \bar{\eta}_\zeta \mathcal{A}(\bar{p}_1, \dots \bar{p}_n)\,.
\eeq
Terms on the left hand side take the general form
\beq \label{Terms}
\left. R^{-\frac{n}{2}}_\zeta \left(  \prod_{i=1}^n  \int_{x_i}   e^{{\rm i} p_i x_i}  \right)    ( -\Box_{x_j} - m^2)    \frac{\delta^m \Psi_{\zeta }^{x_j}\left[ \bar{\phi} \right]}{\delta \phi^{y_1} \dots  \delta \phi^{y_m}}    G^{y_1z_1} \dots  G^{y_m z_m}     \left(\frac{\delta \mathcal{A}}{\delta \psi^{z_1}}  \dots \frac{\delta \mathcal{A}}{\delta \psi^{z_m}}\right)_{; x_1 \dots x_{j-1} x_{j+1} \dots x_n }  \right|_{\psi = 0,p_i\to \bar{p}_i }
\eeq
where $1\leq m \leq n$ and $1\leq j \leq n$. The notation $;x_1 \dots x_{j-1} x_{j+1} \dots x_n$ indicates derivatives with respect to $\psi$ (at those points) where each derivative acts an on a {\it single} $\mathcal{A}$ inside the bracket; one gets terms where the derivatives are distributed in all possible ways.
 These terms will vanish due to the fact that $( -\Box_{x_j} - m^2)   e^{i x_j \bar{p}_j} =  ( \bar{p}^2_j - m^2) e^{i x_j \bar{p}_j}\to  0$ unless there is a compensating pole. Using the pole free condition \eq{pole_free} we can see that there is no compensating pole for $m>1$.
 To understand this better, we can express the terms in momentum space  diagrammatically as in fig.~\ref{General_Diagram} and explained in the caption. In the diagram $m>1$ corresponds to there being more than one $\mathcal{A}$ vertex. In this case the momentum flowing through each propagator is off shell $q^2 \neq m^2$  and the factor $p^2_j-m^2$ is not canceled so the diagram vanishes as the momentum goes on-shell. 
However, for the case $m=1$ one has to take care since the momentum flowing through the single propagator is evaluated at its pole which cancels the corresponding factor due to the inverse propagator as indicated in fig.~\ref{non-zero-diagram}. These terms will produce the terms on the rhs of \eq{Identity}, as is indicated in the diagram. Together, they are given by 
\beq \label{non-zero}
- R^{-n/2}_\zeta \left. \prod_{i=1}^n \int_{x_i}   e^{{\rm i} p_i x_i}  \sum_{j=1}^n  \frac{ \delta^n \mathcal{A}_\zeta[\psi] }{\delta \psi^{x_1} \dots  \delta\psi^{j-1}\delta \psi^z \delta\psi^{j+1} \dots  \delta \psi^{x_n} }  G^{zy}  ( \Box_{x_j} + m^2)    \frac{\delta \Psi_{\zeta }^{x_j}\left[ \bar{\phi} \right]}{\delta \phi^{y}}  \right|_{\psi = 0, p_j \to \bar{p}_j}\,.
\eeq
In position space the cancelation of the propagator and its inverse is seen by using  \eq{eta_zeta}  and \eq{G} which imply (c.f. fig.~\ref{non-zero-diagram})
\beq \label{GBox}
-G^{zy}  (\Box_{x_j} + m^2)    \frac{\delta \Psi_{\zeta }^{x_j}}{\delta \phi^{y}}\left[ \bar{\phi} \right] = \frac{1}{2} \eta_\zeta(-\Box_{x_j})\delta(x_i - z) \,.
\eeq
Finally using $\eta_\zeta(-\Box_{x_j})  e^{i x_j \bar{p}_j} = \bar{\eta}_\zeta e^{i x_j \bar{p}_j} $ we see that \eq{non-zero} is equal to the rhs of \eq{Identity}.

\section{Contrasting field transformations}
Our results suggest that effective actions should be understood as being members of equivalence classes actions where each member of the class is related by a field transformation. Then physics changes as we pass from one equivalence class to another. Further understanding comes from think of  actions as being points in theory space, with the coordinates of theory space being locally mapped to the space of couplings. Locally we can always pick coordinates such that we have inessential coupling $\{\zeta_\a\}$ and essential couplings $\{ \lambda_a\}$. A change in an inessential coupling moves within an equivalence class. A change in an essential coupling moves from one class to another.
In particular, the derivative of the effective action with respect to an inessential coupling is given by a {\it redundant operator} \cite{Wegner:1974sla,Weinberg:1980gg}
\beq \label{eq:zeta_alpha}
\frac{\partial }{\partial \zeta_\a} \Gamma  = - \int_x \Psi_\alpha(x) \frac{\delta \Gamma}{\delta \phi(x)}\,,
\eeq
for some $\Psi_\alpha(x)$.
 The essential operators 
 \beq
 \frac{\partial }{\partial \lambda_a} \Gamma = O_a
 \eeq
are by definition linearly independent of all redundant operators \cite{Wegner:1974sla}. 
Observables only depend on essential couplings such that any observable satisfies \cite{Baldazzi:2021ydj}
\beq
\frac{\partial }{\partial \zeta_\a} \mathcal{O} = 0 \,.
\eeq

A rather different, {\it physics informed}, perspective has been put forward in \cite{Ihssen:2024ihp,Ihssen:2025cff,Bonanno:2025mon}. There the idea is that physics information is
not contained in the effective action alone but in the pair consisting of the effective action and the field transformation.
In other words, given an effective action, one must also know the field transformation to recover the physics content of the
theory. In the most extreme case \cite{Ihssen:2024ihp} advocates fixing the effective action to be equal to the bare one:
\beq \label{Classical_Gamma}
\Gamma_\pp[\phi] = S[\phi] + {\rm constant}
\eeq	
which they refer to as the {\it classical target action}. The idea is then that one can solve 
the QFT instead for the transformation which contains all quantum corrections.
This can be done in practice using functional renormalisation group where the flowing effective action is constrained to equal the bare action at each RG scale  \cite{Ihssen:2024ihp}. Here we discuss  \eq{Classical_Gamma} which is the end point of the flow.
 However, if we fix the full effective action this also seems to fix the essential couplings. Indeed, our correspondence, along with \eq{Classical_Gamma}, implies that 
 \beq \label{Classical_Gamma_trans}
 \Gamma[\phi] = S[\Phi[\phi]] + {\rm constant}
 \eeq
 for some $\Phi[\phi]$. {\it If} the  equivalence theorem were to hold, this would imply that amplitudes gain no quantum corrections. 
 Thus there is some tension between the our perspective developed here and the one that proposes \eq{Classical_Gamma}.
 
 These two points of view can be reconciled if the set of considered field redefinitions, that allow for example \eq{Classical_Gamma_trans} to hold,  is taken to be more general.  In particular, if the assumption \eq{pole_free} is not validated the equivalence theorem would break down and to reconstruct the amplitude  would require knowledge of the field transformation. This may be related to the invertibility of the renormalisation group where information is lost going from the fine grained microscopic description to the coarse grained macroscopic one \cite{Kadanoff:1966wm,Wilson:1973jj}.
 Indeed, the renormalisation group itself can be formulated as a series of changes of variables where an incomplete integration is being performed over high momentum modes. Once the full integral has been completed  all modes are fully integrated out at the end point of the RG flow.

 We can explore the differences in the allowed transformations at one-loop where we have explicitly found the field redefinitions of our correspondence. Writing $\Phi =\phi + \hbar a$, at one-loop \eq{Classical_Gamma_trans} gives 
 \beq \label{Classical_Gamma_trans_one_loop}
 \frac{1}{2} \Tr \log S^{(2)}[\phi] = \frac{\delta S}{\delta \phi^x} a^x  + {\rm constant}\,.
 \eeq
 Then it appears that the only way that we can succeed in solving for $a^x$ is if every operator in  $\frac{1}{2} \Tr \log S^{(2)}[\phi]$ vanishes on the equations of motion, apart from the vacuum term. The alternative would be if $a^x$ is singular on the equations of motion. 
 Nonetheless, we can solve for $a^x$ with some assumptions, if we suppose $S[\phi] = S[-\phi]$ then
  \beq
  \frac{\delta S}{\delta \phi^x} = \phi^y  L_{yx} \,,
  \eeq
  for some $L_{yx}$, and that
 \beq
  \frac{1}{2} \Tr \log S^{(2)}[\phi]  = M_x \phi^x +  {\rm const}\,,
 \eeq
 for some $M_x$. Then we can solve for $a^x$ finding, if $L_{yx}$ is invertible, that
 \beq
 a^x = (L^{-1})^{xy} M_{y}\,.
 \eeq
 Taking $S = \int_x \left[ \frac{1}{2} \partial_\mu \phi    \partial_\mu \phi  + U(\rho)\right]$ where $\rho = \phi^2/2$,
 we have 
 \beq
 L_{xy} = (-\partial^2 + U'(\rho)) \delta(x-y)
 \eeq
 Thus, with some restrictions, we can realise \eq{Classical_Gamma_trans} at one-loop.
The transformation is also seen to be the type that is singular when we go on-shell and thus fits in the loop hole of the equivalence theorem. 
Thus to compute amplitudes we need knowledge of the change of variables. The advantage of this approach is that it may be easier to solve for this map in certain situations.

Here we are more concerned with local transformations that make the distinction between essential and inessential 
couplings meaningful. Then, at least for some observables, no further knowledge of the genuine field transformation is needed, and one can freely make 
effective transformations.  A way to achieve this in practice has been put forward in \cite{Baldazzi:2021ydj}. There the key idea is to fix the inessential 
couplings at the Gaussian fixed point of the theory,  $S = \int_x  \frac{1}{2} \partial_\mu \phi    \partial_\mu \phi$, considering local transformations. 
The essential/inessential distinction is then well defined and, furthermore, extends away from the Gaussian fixed point beyond perturbation theory. Indeed,  if $O_{\a}$ ($O_a$) is the set of redundant operators (essential) {\it at the Gaussian fixed point} we can define a matrix $\Upsilon_{\a\b}$ by
\beq
 - \int_x \Psi_\alpha(x) \frac{\delta \Gamma}{\delta \phi(x)} =  \sum_\beta \Upsilon_{\a\b}  O_{\b} + \sum_a  y_{\a a} O_a\,.
\eeq
where $\{\Psi_\alpha(x)\}$ are a basis of local operators. 
Then when $\Upsilon_{\a\b}$ is invertible the operators $O_a$ remain linearly independent of the redundant operators away from the Gaussian fixed point, and hence remain essential. The {\it minimal essential scheme} which assumes $ \Upsilon_{\a\b}$ is invertible, is naturally restricted to avoid theories with more degrees of freedom such as higher derivative theories.
It also neglects non-local transformations such as those that allow for \eq{Classical_Gamma_trans_one_loop} to be solved for $a^x$.
Applied to critical phenomena it allows one to take full advantage of the invariance properties of the renormalisation group \cite{Wegner:1974sla,JonaLasinio,Green1977InvarianceOC}.

However, it is always the case that, if we observe the field $\cc$ experimentally, we must also find the $\phi[\chi]$ such  $\Gamma_\cc[\chi] = \Gamma_\pp[\phi[\chi]]$ in order  to generate the observed correlation functions from $\Gamma_\pp[\phi]$. In this sense we do need knowledge of the field transformation if we observe correlation functions directly. This is just a special case of the statement that any model requires a physical interpretation in terms of measured quantities. Thus the real challenge is not to find the map between two sets of theoretical variables, but the map from one's chosen set of variables and the correlation functions that are measured in practice.

 \section{Discussion}
In quantum field theory there is a large amount of freedom to pick which variables we work with. Exploiting this we can find the 
most useful variables to understand different physical phenomena. The effective action appears to give a description of physics 
in terms of correlation functions of a specific choice of variable which we couple to the source. Our correspondence suggests that the 
actions which are obtained for different source terms will be related by an effective change of variables of the mean field. 
Furthermore, going  the other way around, we can obtain different actions by making an explicit change of variables at the level 
effective action directly. In this case the correspondence implies that the transformed action generates correlation functions of some composite field variable.
 
 The correspondence between the two types of transformations is at the heart of the Essential Renormalisation Group \cite{Baldazzi:2021ydj} which can be used to study non-perturbative phenomena such as phase transitions and quantum gravity. In that approach a cutoff scale dependent version \cite{Gies:2001nw,Pawlowski:2005xe,Floerchinger:2009uf,Ihssen:2022xjv,Ihssen:2023nqd} of \eq{hat_cov} is performed implicitly allowing one to work only with the essential couplings which are invariant under a reparameterisation of the field variables. 
Such a scheme can be contrast with schemes where one must follow also the flow of the inessential couplings i.e. those couplings whose variation can be absorbed by a change of variables.
Our results suggest that using different schemes ultimately leads to effective actions which are related via \eq{GammaequalsGamma}. This is an important result since it tells us how to compare different effective actions computed using different renormalisation schemes.

Our correspondence rests on the existence of unique solutions initial value problems. The equations can be expanded to any finite loop order to obtain  analytic flows provided the action and transformation, which are known, are analytic. Assuming they are,   
 unique solutions to the initial value problems will exist to each order in perturbation theory.

There is some danger in assuming the correspondence holds non-perturbatively. However, one can compare results obtained assuming the correspondence holds with those obtained without this assumption. Qualitative and quantitive agreement would then supply evidence for the correspondence on a case by case basis.
For example, the results of \cite{Baldazzi:2021ydj}  provide such non-perturbative evidence for the correspondence in the case of the Wilson-Fisher fixed point.

For quantum gauge theories, such as QCD and quantum gravity, classical gauge invariance is  often replaced by the freedom to fix the gauge. It has been pointed out in  \cite{Falls:2025tid} that while gauge independence can be realised   by utilising field reparameterisations,  in the continuum limit the transformations that relate different gauges will be singular. Our findings here suggest that one can equivalently make  non-singular effective transformations to relate effective actions computed in different gauges.
 We leave a full exploration of this idea to future work. 

\section*{Acknowledgements}
I thank David Sutherland and Nicol\'as Wschebor  for their comments and carefully reading the manuscript.
I also thank Jan Pawlowski for his critical remarks on the first draft of this paper.

\begin{appendix}
 \section{Functional identity}
 \label{App:1}
 In this appendix we demonstrate the identity \eq{Functional_identity}. 
 Here for clarity we suppress $\zeta$.
 
 First from the expression for the effective action we have that 
 \beq
\langle \hat{F}[\pp] \rangle[\phi] = {\rm e}^{ \Gamma[\phi] - \phi \cdot  \frac{\delta \Gamma[\phi]}{\delta \phi}  } \hat{F}\left[\frac{1}{\Gamma^{(2)}[\phi]} \cdot \frac{\delta}{\delta \phi} \right] {\rm e}^{- \Gamma[\phi] + \phi \cdot  \frac{\delta \Gamma[\phi]}{\delta \phi} }\,.
\eeq
 Then we note that the $\phi$-derivative will act on the exponential or it will act elsewhere.
 Whenever the derivative acts on the exponential it will have the effect to place $\phi$ in the argument of  $\hat{F}$ since 
  \beq
  \frac{1}{\Gamma^{(2)}[\phi]} \cdot \frac{\delta}{\delta \phi} \left(  - \Gamma[\phi] + \phi \cdot  \frac{\delta \Gamma[\phi]}{\delta \phi}  \right) = \phi\,,
  \eeq
 hence
 \beq
  {\rm e}^{ \Gamma[\phi] - \phi \cdot  \frac{\delta \Gamma[\phi]}{\delta \phi}  } \hat{F}\left[\frac{1}{\Gamma^{(2)}[\phi]} \cdot \frac{\delta}{\delta \phi} \right] {\rm e}^{- \Gamma[\phi] + \phi \cdot  \frac{\delta \Gamma[\phi]}{\delta \phi} } =  \hat{F}\left[ \phi + \frac{1}{\Gamma^{(2)}[\phi]}  \cdot \frac{\delta}{\delta \phi} \right]\,.
 \eeq
Another way to see this is to call
\beq
\mathcal{X} = \frac{1}{\Gamma^{(2)}[\phi]} \cdot \frac{\delta}{\delta \phi}
\eeq
 and 
 \beq
 \mathcal{Y} =  {\rm e}^{- \Gamma[\phi] + \phi \cdot  \frac{\delta \Gamma[\phi]}{\delta \phi} }
 \eeq
then
\beq
[\mathcal{X}  , \mathcal{Y} ] = \phi  \mathcal{Y}  
\eeq
 and
 \beq
  [\mathcal{Y} , \phi] = 0
 \eeq
 which imply
\beq
 \mathcal{Y}^{-1} \mathcal{X}   \mathcal{Y} =  \mathcal{X} + \phi  
\eeq
Then we have that
\beq
 \mathcal{Y}^{-1} \hat{F}[\mathcal{X}] \mathcal{Y} 
 = \hat{F}[  \mathcal{Y}^{-1}  \mathcal{X}  \mathcal{Y} ] =   \hat{F}[\mathcal{X} + \phi ] 
\eeq 
 which is demonstrates the identity \eq{Functional_identity}.
 The identity
 \beq
  e^{S[\pp]}  \tilde{\Psi}\left[- \frac{\delta}{\delta \hat{\phi}} \right] e^{-S[\pp]} =  \tilde{\Psi}\left[ S^{(1)}[\pp]  - \frac{\delta}{\delta \hat{\phi}} \right] 
 \eeq
 which is needed to pass from \eq{Psi_hat} to \eq{Psi_hat2} can be proved similarly. With 
 \beq
 \mathcal{W} = - \frac{\delta}{\delta \hat{\phi}}
 \eeq
 we have
 \beq
  e^{S[\pp]}   \mathcal{W}   e^{-S[\pp]}  =  S^{(1)}[\pp] +  \mathcal{W} 
 \eeq
 and therefore 
 \beq
  e^{S[\pp]}  \tilde{\Psi}\left[  \mathcal{W} \right] e^{-S[\pp]} =   \tilde{\Psi}\left[ e^{S[\pp]}  \mathcal{W}  e^{-S[\pp]} \right]  = \tilde{\Psi}\left[ S^{(1)}[\pp] + \mathcal{W} \right] 
 \eeq
  \end{appendix}


\begin{thebibliography}{99}
 
 %\cite{Cohen:2022uuw}
\bibitem{Cohen:2022uuw}
T.~Cohen, N.~Craig, X.~Lu and D.~Sutherland,
``On-Shell Covariance of Quantum Field Theory Amplitudes,''
Phys. Rev. Lett. \textbf{130} (2023) no.4, 041603
%doi:10.1103/PhysRevLett.130.041603
[arXiv:2202.06965 [hep-th]].
%25 citations counted in INSPIRE as of 04 Jun 2024
 
 
%\cite{Cohen:2023ekv}
\bibitem{Cohen:2023ekv}
T.~Cohen, X.~Lu and D.~Sutherland,
``On amplitudes and field redefinitions,''
JHEP \textbf{06} (2024), 149
%doi:10.1007/JHEP06(2024)149
[arXiv:2312.06748 [hep-th]].
%9 citations counted in INSPIRE as of 04 Mar 2025


%\cite{Chisholm:1961tha}
\bibitem{Chisholm:1961tha}
J.~S.~R.~Chisholm,
``Change of variables in quantum field theories,''
Nucl. Phys. \textbf{26} (1961) no.3, 469-479
doi:10.1016/0029-5582(61)90106-7
%165 citations counted in INSPIRE as of 03 Jun 2024

%\cite{Kamefuchi:1961sb}
\bibitem{Kamefuchi:1961sb}
S.~Kamefuchi, L.~O'Raifeartaigh and A.~Salam,
%``Change of variables and equivalence theorems in quantum field theories,''
Nucl. Phys. \textbf{28} (1961), 529-549
doi:10.1016/0029-5582(61)90056-6
%303 citations counted in INSPIRE as of 03 Jun 2024




%\cite{Efimov:1972juh}
\bibitem{Efimov:1972juh}
G.~V.~Efimov and M.~L.~Rutenberg,
``On the equivalence theorem in quantum field theory,''
Theor. Math. Phys. \textbf{16} (1973) no.2, 763-770
doi:10.1007/BF01037128
%4 citations counted in INSPIRE as of 03 Jun 2024



%\cite{Kallosh:1972ap}
\bibitem{Kallosh:1972ap}
R.~E.~Kallosh and I.~V.~Tyutin,
``The Equivalence theorem and gauge invariance in renormalizable theories,''
Yad. Fiz. \textbf{17} (1973), 190-209
%162 citations counted in INSPIRE as of 11 Jun 2024

%\cite{Arzt:1993gz}
\bibitem{Arzt:1993gz}
C.~Arzt,
``Reduced effective Lagrangians,''
Phys. Lett. B \textbf{342} (1995), 189-195
doi:10.1016/0370-2693(94)01419-D
[arXiv:hep-ph/9304230 [hep-ph]].
%259 citations counted in INSPIRE as of 03 Jun 2024


%\cite{Tyutin:2000ht}
\bibitem{Tyutin:2000ht}
I.~V.~Tyutin,
``Once again on the equivalence theorem,''
Phys. Atom. Nucl. \textbf{65} (2002), 194-202
%doi:10.1134/1.1446571
[arXiv:hep-th/0001050 [hep-th]].
%35 citations counted in INSPIRE as of 11 Jun 2024


%\cite{Falls:2018olk}
\bibitem{Falls:2018olk}
K.~Falls and M.~Herrero-Valea,
``Frame (In)equivalence in Quantum Field Theory and Cosmology,''
Eur. Phys. J. C \textbf{79} (2019) no.7, 595
%doi:10.1140/epjc/s10052-019-7070-3
[arXiv:1812.08187 [hep-th]].
%43 citations counted in INSPIRE as of 14 Apr 2025


%\cite{Falls:2025tid}
\bibitem{Falls:2025tid}
K.~Falls,
``Gauge invariant effective actions for dressed fields,''
[arXiv:2503.05869 [hep-th]].
%2 citations counted in INSPIRE as of 14 Apr 2025



%\cite{Wetterich:2024uub}
\bibitem{Wetterich:2024uub}
C.~Wetterich,
``Field transformations in functional integral, effective action and functional flow equations,''
Nucl. Phys. B \textbf{1008} (2024), 116707
%doi:10.1016/j.nuclphysb.2024.116707
[arXiv:2402.04679 [hep-th]].
%6 citations counted in INSPIRE as of 08 Mar 2025


%\cite{Manohar:2024xbh}
\bibitem{Manohar:2024xbh}
A.~V.~Manohar, J.~Pag\`es and J.~Roosmale Nepveu,
``Field redefinitions and infinite field anomalous dimensions,''
JHEP \textbf{05} (2024), 018
%doi:10.1007/JHEP05(2024)018
[arXiv:2402.08715 [hep-ph]].
%4 citations counted in INSPIRE as of 04 Mar 2025

%\cite{Polchinski:1983gv}
\bibitem{Polchinski:1983gv}
J.~Polchinski,
``Renormalization and Effective Lagrangians,''
Nucl. Phys. B \textbf{231} (1984), 269-295
%doi:10.1016/0550-3213(84)90287-6
%1413 citations counted in INSPIRE as of 18 Aug 2024

%\cite{Wegner:1974sla}
\bibitem{Wegner:1974sla}
F.~J.~Wegner,
``Some invariance properties of the renormalization group,''
J. Phys. C \textbf{7} (1974) no.12, 2098.
%doi:10.1088/0022-3719/7/12/004
%103 citations counted in INSPIRE as of 13 Oct 2024


%\cite{Weinberg:1980gg}
\bibitem{Weinberg:1980gg}
S.~Weinberg,
``Ultraviolet divergences in quantum theories of gravitation,''
in General Relativity: An Einstein Centenary Survey, 1980.
%42 citations counted in INSPIRE as of 13 Sep 2024


%\cite{Pawlowski:2005xe}
\bibitem{Pawlowski:2005xe}
J.~M.~Pawlowski,
``Aspects of the functional renormalisation group,''
Annals Phys. \textbf{322} (2007), 2831-2915
%doi:10.1016/j.aop.2007.01.007
[arXiv:hep-th/0512261 [hep-th]].
%916 citations counted in INSPIRE as of 16 Jun 2024



%\cite{Cohen:2024fak}
\bibitem{Cohen:2024fak}
T.~Cohen, M.~Forslund and A.~Helset,
``Field Redefinitions Can Be Nonlocal,''
[arXiv:2412.12247 [hep-th]].
%1 citations counted in INSPIRE as of 16 Mar 2025

%\cite{Baldazzi:2021ydj}
\bibitem{Baldazzi:2021ydj}
A.~Baldazzi, R.~B.~A.~Zinati and K.~Falls,
``Essential renormalisation group,''
SciPost Phys. \textbf{13} (2022) no.4, 085
%doi:10.21468/SciPostPhys.13.4.085
[arXiv:2105.11482 [hep-th]].
%33 citations counted in INSPIRE as of 16 Jun 2024

%\cite{Ihssen:2024ihp}
\bibitem{Ihssen:2024ihp}
F.~Ihssen and J.~M.~Pawlowski,
``Physics-informed renormalisation group flows,''
[arXiv:2409.13679 [hep-th]].
%10 citations counted in INSPIRE as of 14 Apr 2025

%\cite{Ihssen:2025cff}
\bibitem{Ihssen:2025cff}
F.~Ihssen and J.~M.~Pawlowski,
``Physics-informed gauge theories,''
[arXiv:2503.22638 [hep-th]].
%2 citations counted in INSPIRE as of 28 Apr 2025


%\cite{Bonanno:2025mon}
\bibitem{Bonanno:2025mon}
A.~Bonanno, F.~Ihssen and J.~M.~Pawlowski,
``Tunneling with physics-informed RG flows in the anharmonic oscillator,''
[arXiv:2504.03437 [hep-th]].
%0 citations counted in INSPIRE as of 28 Apr 2025


%\cite{Kadanoff:1966wm}
\bibitem{Kadanoff:1966wm}
L.~P.~Kadanoff,
``Scaling laws for Ising models near T(c),''
Physics Physique Fizika \textbf{2} (1966), 263-272
%doi:10.1103/PhysicsPhysiqueFizika.2.263
%543 citations counted in INSPIRE as of 03 May 2025


%\cite{Wilson:1973jj}
\bibitem{Wilson:1973jj}
K.~G.~Wilson and J.~B.~Kogut,
``The Renormalization group and the epsilon expansion,''
Phys. Rept. \textbf{12} (1974), 75-199
%doi:10.1016/0370-1573(74)90023-4
%3401 citations counted in INSPIRE as of 03 May 2025


\bibitem{JonaLasinio}
G.~Jona-Lasinio
``Generalized Renormalization Transformations,''
in Proc. Nobel Symp. 24: Collective Properties of Physical Systems, Stockholm, Nobel Foundation; New York, Academic Press,
Edited by Svartholm, Nils,
 ISBN: 0-12-460350-5
 
 \bibitem{Green1977InvarianceOC}
M.~S.~Green
``Invariance of critical exponents for renormalization groups generated by a flow vector'',
Physical Review B, 1977, 15


%\cite{Gies:2001nw}
\bibitem{Gies:2001nw}
H.~Gies and C.~Wetterich,
``Renormalization flow of bound states,''
Phys. Rev. D \textbf{65} (2002), 065001
%doi:10.1103/PhysRevD.65.065001
[arXiv:hep-th/0107221 [hep-th]].
%242 citations counted in INSPIRE as of 09 Mar 2025


%\cite{Floerchinger:2009uf}
\bibitem{Floerchinger:2009uf}
S.~Floerchinger and C.~Wetterich,
``Exact flow equation for composite operators,''
Phys. Lett. B \textbf{680} (2009), 371-376
%doi:10.1016/j.physletb.2009.09.014
[arXiv:0905.0915 [hep-th]].
%125 citations counted in INSPIRE as of 09 Mar 2025


%\cite{Ihssen:2022xjv}
\bibitem{Ihssen:2022xjv}
F.~Ihssen and J.~M.~Pawlowski,
``Functional flows for complex effective actions,''
SciPost Phys. \textbf{15} (2023) no.2, 074
%doi:10.21468/SciPostPhys.15.2.074
[arXiv:2207.10057 [hep-th]].
%18 citations counted in INSPIRE as of 28 Apr 2025


%\cite{Ihssen:2023nqd}
\bibitem{Ihssen:2023nqd}
F.~Ihssen and J.~M.~Pawlowski,
``Flowing fields and optimal RG-flows,''
[arXiv:2305.00816 [hep-th]].
%14 citations counted in INSPIRE as of 28 Apr 2025



 \end{thebibliography}
\end{document}